\documentclass[english,aip,jcp,reprint,showpacs,superscriptaddress]{revtex4-1}

\usepackage{units}
\usepackage{amsmath}
\usepackage{amssymb}
\usepackage{graphicx}
\usepackage{esint}
\usepackage{color}
\usepackage{units}
\usepackage{babel}

\begin{document}

\title{Dynamic principle for ensemble control tools}

\author{A. Samoletov}
\email{A.Samoletov@liverpool.ac.uk}

\affiliation{Department of Mathematical Sciences, University of Liverpool, Liverpool,
UK}

\affiliation{Institute for Physics and Technology, Donetsk, Ukraine}

\author{B. Vasiev}
\email{B.Vasiev@liverpool.ac.uk}

\affiliation{Department of Mathematical Sciences, University of Liverpool, Liverpool,
UK}

\date{ } 
\begin{abstract}
Dynamical equations describing physical systems in contact with the
thermal bath are commonly extended by mathematical tools called ``thermostats''. These tools are designed for sampling ensembles in statistical mechanics.
Here we propose a dynamic principle underlying a range of thermostats
which is derived using fundamental laws of statistical physics and
insures invariance of the canonical measure. 
The principle covers both stochastic and deterministic thermostat schemes.
 Our method has a clear advantage over a range of proposed and widely used thermostat schemes which are based on formal mathematical reasoning. Following the derivation of proposed principle we show its generality and illustrate its  
applications including design of temperature
control tools that differ from the Nos{\'e}-Hoover-Langevin scheme. 
\end{abstract}
\maketitle

\section{Introduction}

Analysis of molecular systems is an essential part of research in
a range of disciplines in natural sciences and in engineering \cite{AllenTildesley1989,frenkel2002understanding}.
As molecular systems affected by environmental thermodynamic conditions,
they are studied in the context of statistical physics ensembles.
Methods of dynamical sampling of the corresponding probability measures
are important for applications and they are under extensive study
and development \cite{LeimkuhlerMatthews2015,tuckerman2010statistical,JeppsRondoni2010,BussiDonadioParrinello2007,SamoletovDettmannChaplain2007,LeimkuhlerNoorizadehTheil2009,SamoletovDettmannChaplain2010,Leimkuhler2010,BajarsFrankLeimkuhler2011,DiPierroElberLeimkuhler2015}.
The traditional application of thermostats is molecular dynamics (MD),
that is sampling of equilibrium systems with known potential energy
functions, $V(q)$, where $q$ is a system\color{red}'s \color{black} configuration. However,
the ability to sample equilibrium ensembles at constant temperature
$T$ would also imply the ability to sample arbitrary probability
measures. Indeed, as an alternative to the conventional MD practice,
one may use a probability density $\sigma(q)$, theoretical or extracted
from experimental data, to define the potential function as $V(q)=-k_{B}T\,\ln\sigma(q)$,
where $k_{B}$ is the Boltzmann constant.

Thermostats embedded into dynamical equations bring in the so-obtained
dynamics rich mathematical content. Such dynamical systems with an
invariant probability measure have become increasingly popular for
mathematical studies in a wide range of applications including investigation
of non-equilibrium phenomena \cite{JeppsRondoni2010,Dittmar2014,ness2016nonequilibrium,dittmar2016ordering,Stella2014,LepriLiviPoliti1997,pastorino2007comparison,ciccotti2016non},
mathematical biology models \cite{bianca2012thermostatted,SamoletovVasiev2013,chow2005emotion,tang2016non},
multiscale models \cite{MonesJonesGoetzEtAl2015,fritz2011multiscale,praprotnik2008multiscale,Chen2016,Leimkuhler2016},
Bayesian statistics and Bayesian machine learning applications \cite{Leimkuhler2016,Ding2014,NIPS2015_5978,noid2013perspective},
superstatistics \cite{Fukuda2017,FukudaMoritsugu2015}.

Here, we present a unified approach for derivation of thermostats
sampling the canonical ensemble. The corresponding method is derived using fundamental
physical arguments that facilitate understanding physics of thermostat
schemes in general, and elucidate physics of the Nos{\'e}-Hoover (NH)
and the Nos{\'e}-Hoover-Langevin (NHL) dynamics in particular. Besides,
our method allows to build a plethora of thermostats, stochastic as
well as deterministic, including those previously proposed. We expect
that it can also be adjusted to arbitrary probability measures.

Classical mechanics and equilibrium statistical physics are adequately
described in terms of the Hamiltonian dynamics. Dynamic thermostat
schemes involve modified Hamiltonian equations of motion where certain
temperature control tools are included. The modified dynamics can
be deterministic as well as stochastic \cite{LeimkuhlerMatthews2015,tuckerman2010statistical,SamoletovDettmannChaplain2007,LeimkuhlerNoorizadehTheil2009,SamoletovDettmannChaplain2010,Leimkuhler2010,LeimkuhlerMatthews2013,AllenTildesley1989,hoover2012computational,BussiDonadioParrinello2007,SamoletovVasiev2013,hoover1985canonical,nose1984molecular,frenkel2002understanding,BajarsFrankLeimkuhler2011,JeppsRondoni2010,BussiZykova-TimanParrinello2009,DiPierroElberLeimkuhler2015}.
Recently proposed NHL thermostats \cite{LeimkuhlerMatthews2015,SamoletovDettmannChaplain2007,LeimkuhlerNoorizadehTheil2009,SamoletovDettmannChaplain2010}
combine deterministic dynamics with stochastic perturbations. This
combination ensures ergodicity and allows ``gentle'' perturbation
of the physical dynamics that is often desired \cite{LeimkuhlerMatthews2015,LeimkuhlerNoorizadehTheil2009}.

To introduce our scheme, we consider a dynamical system $\mathrm{S}$
consisting of $N$ particles in $d$-dimensional space ($\mathcal{N}=dN$
degrees of freedom) described by the Hamiltonian function $H(x)$,
where $x=(p,q)$ is a point in the phase space $\mathcal{M=\mathbb{R}}^{2dN}$,
$p=\left\{ \mathbf{p}_{i}\in\mathbb{R}^{d}\right\} _{i=1}^{N}$ are
momentum variables and $q=\left\{ \mathbf{q}_{i}\in\mathbb{R}^{d}\right\} _{i=1}^{N}$
are position variables. The Hamiltonian dynamics has the form, $\dot{x}=\boldsymbol{J}\boldsymbol{\nabla}H(x)$
in the phase space $\mathcal{M}$, where ${\normalcolor \boldsymbol{J}}$
is the symplectic unit. The canonical ensemble describes the system
$\mathrm{S}$ in contact with the heat bath $\Sigma$ (an energy reservoir
permanently staying in the thermal equilibrium with the thermodynamic
temperature $T$), and $\mathrm{S}$ may exchange energy with $\Sigma$
only in the form of heat. Thus, the temperature of the system $\mathrm{S}$
is fixed while its energy, $E$, is allowed to fluctuate. The canonical
distribution has the form $\rho_{\infty}(x)\propto\exp\left[-\beta H\left(x\right)\right]$,
where $\beta=(k_{B}T)^{-1}$. On average along an ergodic trajectory
$\left\langle E(t)\right\rangle =E(T)=const$. Rate of energy exchange
between the system $\mathrm{S}$ and the thermal bath $\Sigma$ depends
on the temperature $T$. Note that Hamiltonian system is unable to
sample the canonical distribution since there is no energy exchange
between the system and the heat bath. To describe the heat transfer,
it is necessary to modify the equations of motion in a way that the
dynamics becomes non-Hamiltonian \cite{Ruelle2004}. Suppose $\dot{x}=\boldsymbol{G}(x)$
is a modified law of motion and $\dot{H}(x)=\boldsymbol{G}(x)\cdot\boldsymbol{\nabla}H(x)$
is the rate of energy change (depending on $T$) such that $\left\langle \boldsymbol{G}(x)\cdot\boldsymbol{\nabla}H(x)\right\rangle =0$,
that is the energy is constant on average. Let $\boldsymbol{G}(x)\cdot\boldsymbol{\nabla}H(x)\propto F(x,\beta)$
where the temperature dependence is a key. In order to state the dynamic
principle governing temperature control tools, a few definitions are
required.

\section{Microscopic temperature expressions}

\emph{Consider $F(x,\beta)$ such that $\left\langle F(x,\beta)\right\rangle =0$
for all $\beta>0$. This condition is denoted as $F(x,\beta)\sim0$
while the function $F(x,\beta)$ is called the microscopic temperature
expression (TE).}

For the system with $H(x)=K(p)+V(q)$ examples of TEs include the
kinetic TE, $F_{k\text{in}}(p,\beta)=2K(p)\beta-\mathcal{N}$, and
the configurational TE, $F_{conf}(q,\beta)=(\boldsymbol{\nabla}V(q))^{2}\beta-\Delta V(q)$
\cite{rugh1997dynamical}.

Various TEs can be obtained in the following manner. Suppose that
$F(x,\beta)$ is a polynomial in $\beta$, $F(x,\beta)=\sum_{n=0}^{2L+1}\varphi_{n}(x)\beta^{n}\sim0$,
where $L\in\mathbb{Z}_{\geq0}$ and functions $\left\{ \varphi_{n}(x)\right\} _{n=0}^{2L+1}$,
are subject to specification. Rewrite $F(x,\beta)$ in the form
\begin{equation}
F(x,\beta)=\sum_{k=0}^{L}\left(\varphi_{2k}(x)+\beta\varphi_{2k+1}(x)\right)\beta^{2k}\sim0\label{eq:GenTE1}
\end{equation}
for all $\beta>0$. Thus, from \eqref{eq:GenTE1} it follows that $\varphi_{2k}(x)+\beta\varphi_{2k+1}(x)\sim0$
for all $k\in\left\{ 0,1,\ldots,L\right\} $. 
To find $\varphi_{2k}(x)$ and $\varphi_{2k+1}(x)$ satisfying this condition consider the basic expression, $F(x,\beta)=\beta\,\varphi_{1}(x)+\varphi_{0}(x)$.
Substituting $\varphi(x)\partial_{i}H(x)$ for $\varphi_{1}(x)$,
where $\varphi(x)$ is an arbitrary function, and utilizing the identity,
$\partial_{i}e^{-\beta H(x)}=-\beta\partial_{i}H(x)e^{-\beta H(x)}$
for all $i=1,\ldots,2dN$ and $x\in\mathcal{M}$, where $\partial_{i}\equiv\nicefrac{\partial}{\partial x_{i}}$,
we get $\partial_{i}\varphi(x)+\varphi_{0}(x)\sim0$.
Then, excluding $\varphi_{0}(x)$ from the basic expression, we arrive at $F(x,\beta)=\beta\,\varphi(x)\partial_{i}H(x)-\partial_{i}\varphi(x)\sim0$ (or $\varphi(x)\partial_{i}H(x)-k_{B}T\partial_{i}\varphi(x)\sim0$) for each and every $x_{i}$ in $\mathcal{M}$ provided that   $\varphi(x)\exp[-\beta H(x)]\rightarrow0$
as $\left|x\right|\rightarrow\infty$. This result can be represented in a compact form. Suppose $\boldsymbol{\varphi}_0(x)$
is a vector field on $\mathcal{M}$ such that $\boldsymbol{\varphi}_0(x)\exp[-\beta H(x)]\rightarrow\boldsymbol{0}$
as $\left|x\right|\rightarrow\infty$. Then 
\begin{equation}
F_{0}(x,T)=\boldsymbol{\boldsymbol{\varphi}}_0(x)\cdot\boldsymbol{\nabla}H(x)-k_{B}T\boldsymbol{\nabla}\cdot\boldsymbol{\varphi}_0(x)\sim0.\label{eq:Texpression}
\end{equation}
This form of TE was previously discussed\cite{jepps2000microscopic}. More
general TEs are allowed, \textit{e.g.} vector fields $\boldsymbol{F}(x,\beta)=\beta\,\boldsymbol{\nabla}H(x)\times\boldsymbol{\varphi}(x)-\boldsymbol{\nabla}\times\boldsymbol{\varphi}(x)\sim0$,
and so on. As a further generalization we introduce the notation
\[
F_{l}(x,T)=\boldsymbol{\boldsymbol{\varphi}}_{l}(x)\cdot\boldsymbol{\nabla}H(x)-k_{B}T\boldsymbol{\nabla}\cdot\boldsymbol{\varphi}_{l}(x),
\]
where $l=0,1,...,L$, $\boldsymbol{\boldsymbol{\varphi}}_{0}(x)=\boldsymbol{\boldsymbol{\varphi}}(x)$,
and $\left\{ \boldsymbol{\boldsymbol{\varphi}}_{l}(x)\right\} _{l=0}^{L}$
is a set of vector fields such that $\boldsymbol{\varphi}_{l}(x)\exp[-\beta H(x)]\rightarrow\boldsymbol{0}$
as $\left|x\right|\rightarrow\infty$. Then the general scalar
TE can be represented as
\begin{gather}
F_{L}(x,T)=\sum_{l=0}^{L}F_{l}(x,T)(k_{B}T)^{2l}\sim0\label{eq:GenTE}
\end{gather}
for all $L\in\mathbb{Z}_{\geq0}$. A particular example of the use
of such a TE in a limited context ($L=1$ and $\boldsymbol{\varphi}_{l}(x)\propto(\boldsymbol{p},0)$
leading to the kinetic TE) can be found in the literature\cite{HooverHolian1996}.
In what follows we focus mainly on $F_{0}(x,T)$ and only to a certain extent on $F_{L}(x,T)$ where $\,L\geq1$.

Although the expression \eqref{eq:Texpression} implies the existence
of infinite number of TEs, they all are equivalent from the thermodynamic
perspective. However, the time interval required to achieve a specified
accuracy in $\left\langle F(x,\beta)\right\rangle =0$ can differ
for different TEs \cite{uhlenbeck1963lectures}. In general, physical
systems are often distinguished by multimodal distributions and by
existence of metastable states. Their dynamics is characterized by
processes occurring on a number of timescales. We assume that TEs
can be associated with dynamical processes occurring on various time
scales, and thus, they can be combined in multiscale models.

\section{Dynamic principle}

Now we claim the following dynamic principle for ensemble control
tools: \emph{ Let   $F(x,T)$ be a TE. Then there exists the
dynamical system, $\dot{x}=\boldsymbol{G}(x)$, such that} 
\begin{equation}
\boldsymbol{\nabla}H(x)\cdot\boldsymbol{G}(x)\propto F(x,T).\label{eq:conjecture}
\end{equation}
Relationship \eqref{eq:conjecture} states that the rates of dynamical
fluctuations in energy and in TE are proportional, both are zero on
average and there is no energy release along a whole trajectory in
$\mathcal{M}$. It is a necessary condition for any thermostat. In
what follows, with implication of the fundamental requirements of
statistical physics, we show that the relationship \eqref{eq:conjecture}
leads to a general method for obtaining stochastic and deterministic
thermostats.

Let us consider the exchange of energy between the system $\mathrm{S}$
and the thermal bath $\Sigma$. Any system placed in the heat bath
should in some extent perturb it and be affected by backward influence
of this perturbation. There exists a subsystem $\mathrm{S_{ad}}$
of $\Sigma$ such that $\mathrm{S_{ad}}$ is involved in a joint dynamics
with $\mathrm{S}$. The rest of the heat bath is assumed to be unperturbed,
permanently staying in thermal equilibrium. This is an approximation
that is based on separation of relevant time scales. For instance,
Brownian dynamics assumes that characteristic time scales of $\mathrm{S}$
and $\Sigma$ are well separated and the system $\mathrm{S}$ does
not perturb $\Sigma$. If the time scale is refined (which is of particular
importance for small systems) then we have to take into account joint
dynamics of $\mathrm{S}$ and $\mathrm{S_{ad}}$. We will show that
this case is closely related to NHL \cite{SamoletovDettmannChaplain2007,LeimkuhlerNoorizadehTheil2009}
and NH dynamics \cite{nose1984molecular,hoover1985canonical}.

Thus, we have two cases: (\textsf{\textbf{A}}) the system $\mathrm{S}$
doesn't perturb the thermal bath and there are no new dynamic variables.
The thermal bath in this case can only be taken into account implicitly
via stochastic perturbations (similar to the Langevin dynamics); (\textsf{\textbf{B}})
the system $\mathrm{S}$ perturbs a part ($\mathrm{S_{ad}}$) of the
thermal bath $\Sigma$, while the rest of the thermal bath remains
unperturbed. We assume that there is no
direct energy exchange between $\mathrm{S}$ and $\Sigma$. 
Fundamentals of the statistical mechanics require that
the systems $\mathrm{S}$ and $\mathrm{S_{ad}}$ are statistically
independent at thermal equilibrium. 
Let us consider cases \textsf{\textbf{A}} and\textbf{ }\textsf{\textbf{B}}
in detail.

\subsection{Stochastic dynamics}

Suppose \textbf{ $\:\boldsymbol{\nabla}H(x)\cdot\dot{x}=\lambda F_{0}(x,T)$},
where $\lambda$ is a constant. Without loss of generality, we can
consider modified Hamiltonian dynamics in the form, $\dot{x}=\boldsymbol{J}\boldsymbol{\nabla}H(x)+\boldsymbol{\psi}(x,\lambda)$,
and consequently: 
\begin{gather}
\boldsymbol{\nabla}H(x)\cdot\boldsymbol{\boldsymbol{\psi}}(x,\lambda)=\lambda F_{0}(x,T),\label{eq:psi-stoch}
\end{gather}
where the vector field $\boldsymbol{\boldsymbol{\psi}}(x,\lambda)$
is to be found. Since the thermal bath does not appear in equation
\eqref{eq:psi-stoch} explicitly, only stochastic thermal noise may
be involved in the dynamics. To find $\boldsymbol{\boldsymbol{\psi}}$,
we introduce $2\mathcal{N}$-vector of independent thermal white noises,
$\boldsymbol{\xi}(t)$, such that $\left\langle \boldsymbol{\xi}(t)\right\rangle =\boldsymbol{0}$,
$\quad\left\langle \xi_{i}(t)\xi_{j}(t')\right\rangle =2\lambda k_{\mathrm{B}}T\delta_{ij}\delta(t-t')$,
and the vector field, $\boldsymbol{\Phi}(x)$, such that
\[
\left\langle \boldsymbol{\xi}(t)\cdot\boldsymbol{\Phi}(x)\right\rangle =\lambda k_{\mathrm{B}}T\,\langle\boldsymbol{\nabla}\cdot\boldsymbol{\varphi}(x)\rangle,
\]
where $\left\langle \cdots\right\rangle $ is the Gaussian average
over all realizations of $\boldsymbol{\xi}(t)$. Using Novikov's formula
\cite{novikov1965functionals,klyatskin2005dynamics}, we get
\[
\left\langle \boldsymbol{\xi}(t)\cdot\boldsymbol{\Phi}(x)\right\rangle =\,\sum_{i,k}\left\langle \frac{\partial\Phi_{k}}{\partial x_{i}}\frac{\delta x_{i}(t)}{\delta\xi_{k}(t)}\right\rangle \lambda k_{\mathrm{B}}T.
\]
Suppose $\frac{\delta x_{i}(t)}{\delta\xi_{k}(t)}=\zeta_{i}(x)\delta_{ik}$,
where the vector field $\boldsymbol{\zeta}(x)$ is such that each
component $\zeta_{i}(x)$ does not depend on $x_{i}$, that is
\[
\boldsymbol{\nabla}\circ\boldsymbol{\zeta}(x)=\boldsymbol{0},
\]
where $\circ$ denotes the component-wise (Hadamard) product of two
vectors and $\boldsymbol{0}$ is the null vector. Then $\boldsymbol{\nabla}\cdot\boldsymbol{\varphi}(x)=\boldsymbol{\nabla}\cdot(\boldsymbol{\zeta}(x)\circ\boldsymbol{\Phi}(x))$.
Thus, we get $\boldsymbol{\varphi}(x)=\boldsymbol{\zeta}(x)\circ\boldsymbol{\Phi}(x)$
and it follows that $\boldsymbol{\Phi}(x)=\boldsymbol{\zeta}^{-1}(x)\circ\boldsymbol{\varphi}(x)$,
where $\boldsymbol{\zeta}^{-1}(x)$ is the vector field such that
$\boldsymbol{\zeta}^{-1}(x)\circ\boldsymbol{\zeta}(x)=\boldsymbol{1}$.
Assuming $\boldsymbol{\varphi}(x)=\boldsymbol{\eta}(x)\circ\boldsymbol{\nabla}H(x)$,
where $\boldsymbol{\eta}(x)\equiv\boldsymbol{\zeta}(x)\circ\boldsymbol{\zeta}(x)$,
we get 
\[
\boldsymbol{\psi}(x,\lambda)=-\lambda\boldsymbol{\eta}(x)\circ\boldsymbol{\nabla}H(x)+\boldsymbol{\zeta}(x)\circ\boldsymbol{\xi}(t)
\]
and the modified Hamiltonian dynamics takes the form of stochastic
differential equation (SDE): 
\begin{equation}
\dot{x}=\boldsymbol{J}\boldsymbol{\nabla}H(x)-\lambda\boldsymbol{\eta}(x)\circ\boldsymbol{\nabla}H(x)+\boldsymbol{\zeta}(x)\circ\boldsymbol{\xi}(t).\label{eq:deformed-StochEqM}
\end{equation}

The Fokker-Planck equation (FPE) corresponding to SDE \eqref{eq:deformed-StochEqM}
has the form $\partial_{t}\rho=\boldsymbol{\mathcal{F}}^{*}\rho$,
where 
\begin{gather*}
\boldsymbol{\mathcal{F}}^{*}\rho=-\boldsymbol{J}\boldsymbol{\nabla}H(x)\cdot\boldsymbol{\boldsymbol{\nabla}}\rho+\lambda\boldsymbol{\boldsymbol{\nabla}}\cdot[\boldsymbol{\eta}(x)\circ\boldsymbol{\nabla}H(x)\,\rho]\\
+\lambda k_{B}T\!\boldsymbol{\nabla\cdot}[\boldsymbol{\eta}(x)\circ\boldsymbol{\nabla}\rho].
\end{gather*}
Note that the last term here was found using the following specific relationship for the vector field
$\boldsymbol{\zeta}(x)$:
\[
\left(\boldsymbol{\zeta}(x)\circ\boldsymbol{\nabla}\right)\cdot\left(\boldsymbol{\zeta}(x)\circ\boldsymbol{\nabla}\rho\right)=\boldsymbol{\nabla\cdot}[\boldsymbol{\eta}(x)\circ\boldsymbol{\nabla}\rho].
\]
 
Invariant probability density for dynamics \eqref{eq:deformed-StochEqM}
is determined by the equation $\boldsymbol{\mathcal{F}}^{*}\rho=0$.
It is expected that this is a unique invariant density \cite{mattingly2002ergodicity,LeimkuhlerNoorizadehTheil2009}. 

We claim that \emph{for the defined above vector field $\boldsymbol{\zeta}(y)$
the canonical density, $\rho_{\infty}\propto\exp\left[-\beta H\left(x\right)\right]$,
is invariant for the stochastic dynamics given by \eqref{eq:deformed-StochEqM},
that is $\boldsymbol{\mathcal{F}}^{*}\rho_{\infty}=0$. } The proof
is by direct calculation.

The Langevin equation is a particular case of \eqref{eq:deformed-StochEqM}.
For example, for the system with $H(x)=\nicefrac{p^{2}}{2m}+V(q)$,
where $x=(p,q)\in\mathbb{R}^{2}$ we have:
\begin{itemize}
\item[]  if $\boldsymbol{\zeta}=(1,0)$, then\\ $\dot{p}=-V'(q)-\lambda\nicefrac{p}{m}+\xi(t)$,
	$\dot{q}=\nicefrac{p}{m}$;
\item[]if $\boldsymbol{\zeta}=(0,1)$, then\\ $\dot{p}=-V'(q)$, $\dot{q}=\nicefrac{p}{m}-\lambda V'(q)+\xi(t)$.
\end{itemize}

The procedure for obtaining stochastic dynamics \eqref{eq:deformed-StochEqM} is essentially a general and can be a quite straightforwardly extended to other TEs, for example, the general scalar TE \eqref{eq:GenTE}.
Indeed,
let us introduce the set of $2\mathcal{N}$-vectors of independent
thermal white noises, $\left\{ \boldsymbol{\xi}(l;t)\right\} _{l=0}^{L},\,L\in\mathbb{Z}_{\geq0}$,
such that $\left\langle \boldsymbol{\xi}(l;t)\right\rangle =\boldsymbol{0}$,
$\quad\left\langle \xi_{i}(l;t)\xi_{j}(l';t')\right\rangle =2\lambda_{l}k_{\mathrm{B}}T\delta_{ij}\delta_{ll'}\delta(t-t')$,
and the set of vector fields, $\left\{ \boldsymbol{\zeta}(l;x)\right\} _{l=0}^{L},\,L\in\mathbb{Z}_{\geq0}$,
such that $\boldsymbol{\nabla}\circ\boldsymbol{\zeta}(l;x)=\boldsymbol{0}$
for any $l\geq0$, where $\circ$ denotes the component-wise (Hadamard)
product of two vectors and $\boldsymbol{0}$ is the null vector. Starting
from the relationship,
\[
\boldsymbol{\nabla}H(x)\cdot\boldsymbol{\boldsymbol{\psi}}(x,\lambda)=\sum_{l=0}^{L}\lambda_{l}F_{l}(x,T)\,(k_{B}T)^{2l},
\]
and then strictly following arguments as stated above, we get
\begin{multline*}
\boldsymbol{\psi}(x,\lambda)=-\sum_{l=0}^{L}\lambda_{l}\boldsymbol{\eta}(l;x)\circ\boldsymbol{\nabla}H(x)\,(k_{B}T)^{2l}\\
+\sum_{l=0}^{L}\boldsymbol{\zeta}(l;x)\circ\boldsymbol{\xi}(l;t)\,(k_{B}T)^{l}.
\end{multline*}
 where $\boldsymbol{\eta}(x)\equiv\boldsymbol{\zeta}(x)\circ\boldsymbol{\zeta}(x)$.
Thus, we arrive at the following stochastic dynamics
\begin{multline}
\dot{x}=\boldsymbol{J}\boldsymbol{\nabla}H(x)-\sum_{l=0}^{L}\lambda_{l}\boldsymbol{\eta}(l;x)\circ\boldsymbol{\nabla}H(x)\,(k_{B}T)^{2l}\\
+\sum_{l=0}^{L}\boldsymbol{\zeta}(l;x)\circ\boldsymbol{\xi}(l;t)\,(k_{B}T)^{l}.\label{eq:deformed-StochEqM-Gen}
\end{multline}
One can verify that the canonical measure is invariant for this stochastic
equation of motion. Generally speaking, the dynamics 
\eqref{eq:deformed-StochEqM-Gen}
includes $2\mathcal{N}(L + 1)$ independent white noise processes. This seems impractical. However,
we can point out that \eqref{eq:deformed-StochEqM-Gen} potentially
useful for multi-timescale stochastic simulations. As a simple example,
let $H(x)=\nicefrac{p^{2}}{2m}+V(q)$, $L=1$, $\boldsymbol{\zeta}(0;x)=(1,0)$,
and $\boldsymbol{\zeta}(1;x)=(0,1)$, then we arrive at the stochastic
dynamics with two timescales involved,
\begin{gather*}
\dot{p}=-V'(q)-\lambda_{0}\frac{p}{m}+\xi_{p}(0;t),\\
\dot{q}=\frac{p}{m}-\lambda_{1}\left(k_{B}T\right)^{2}\,V'(q)+k_{B}T\,\xi_{q}(1;t),
\end{gather*}
where $\left\langle \xi_{p}(0;t)\right\rangle =0,\,\left\langle \xi_{q}(1;t)\right\rangle =0$,
$\left\langle \xi_{p}(0;t)\xi_{q}(1;t)\right\rangle =0$, $\left\langle \xi_{p}(0;t)\xi_{p}(0;t')\right\rangle =2\lambda_{0}k_{\mathrm{B}}T\delta(t-t')$,
$\left\langle \xi_{q}(1;t)\xi_{q}(1;t')\right\rangle =2\lambda_{1}k_{\mathrm{B}}T\delta(t-t')$,
as specified above. 
Analysis of $p-$ and $q-$dynamics can be performed in reduced systems following the separation of these variables according to their time scales
\cite{Samoletov1999}.

\subsection{Deterministic and stochastic dynamics}

Let $\mathrm{S_{ad}}$ be associated\textbf{ }with an even-dimensional
phase space $\mathcal{M}_{\mathrm{ad}}$, the Hamiltonian function
$h(y)$, $y\in\mathcal{M}_{\mathrm{ad}}$, and the Hamiltonian dynamics,
$\dot{y}=\boldsymbol{J}_{y}\boldsymbol{\nabla}_{y}h(y)$, where $\boldsymbol{J}_{y}$
is the symplectic unit. Without loss of generality, we can assume
that the modified Hamiltonian dynamics of the system composed by $\mathrm{S}$
and $\mathrm{S_{ad}}$ has the form, 
\begin{align*}
\dot{x}=\boldsymbol{J}_{x}\boldsymbol{\nabla}_{x}H(x)+\boldsymbol{\psi}(x,y),\\
\dot{y}=\boldsymbol{J}_{y}\boldsymbol{\nabla}_{y}h(y)+\boldsymbol{\psi}^{\ast}(y,x),
\end{align*}
where $\boldsymbol{\psi}(x,y)$ and $\boldsymbol{\psi}^{*}(y,x)$
are vector fields on $\mathcal{M}$ and $\mathcal{M}_{\mathrm{ad}}$
correspondingly. To derive deterministic dynamics, let us temporarily
ignore the heat exchange between $\mathrm{S_{ad}}$ and $\Sigma$, that is, $\boldsymbol{\nabla}_{y}h(y)\cdot\boldsymbol{\psi}^{\ast}(y,x)=\lambda^{\ast}{F}^{\ast}(y,T)$
and $\boldsymbol{\nabla}_{x}H(x)\cdot\boldsymbol{\psi}(x,y)=\lambda{F}(x,T)$.
As discussed above, these  relationships lead to the stochastic dynamics.
Systems
$\mathrm{S}$ and $\mathrm{S_{ad}}$ must be statistically independent
in the thermal equilibrium, so that $\boldsymbol{\nabla}_{x}H(x)\cdot\dot{x}\sim0$
and $\boldsymbol{\nabla}_{y}h(y)\cdot\dot{y}\sim0$ are satisfied
simultaneously. Thus, we assume that 
\begin{align*}
\boldsymbol{\nabla}_{x}H(x)\cdot\boldsymbol{\psi}(x,y) & =g(x)F_{0}^{\ast}(y,T),\\
\boldsymbol{\nabla}_{y}h(y)\cdot\boldsymbol{\psi}^{\ast}(y,x) & =-g^{\ast}(y)F_{0}(x,T),
\end{align*}
where $g(x)$ and $g^{\ast}(y)$ are some vague functions, and
\begin{align}
F_{0}(x,T)=\boldsymbol{\boldsymbol{\varphi}}(x)\cdot\boldsymbol{\nabla}_{x}H(x)-k_{B}T\,\boldsymbol{\nabla}_{x}\cdot\boldsymbol{\varphi}(x),\nonumber \\
F_{0}^{\ast}(y,T)=\boldsymbol{Q}(y)\cdot\boldsymbol{\nabla}_{y}h(y)-k_{B}T\,\boldsymbol{\nabla}_{y}\cdot\boldsymbol{Q}(y),\label{eq:TEs-GNH}
\end{align}
are TEs for the systems $\mathrm{S}$ and $\mathrm{S_{ad}}$ correspondingly.
These relationships are valid for any $H(x)$ and $h(y)$. To specify
$\boldsymbol{\psi}(x,y)$ and $\boldsymbol{\psi}^{*}(y,x)$, we assume
that $g(x)=\boldsymbol{\boldsymbol{a}}(x)\cdot\boldsymbol{\nabla}_{x}H(x)$,
$g^{\ast}(y)=\boldsymbol{b}(y)\cdot\boldsymbol{\nabla}_{y}h(y)$,
where $\boldsymbol{a}(x)$ and $\boldsymbol{b}(y)$ are vector fields
on $\mathcal{M}$ and $\mathcal{M}_{\mathrm{ad}}$, respectively.
It follows that
\begin{gather*}
\boldsymbol{\psi}(x,y)=\boldsymbol{\boldsymbol{a}}(x)\,F_{0}^{\ast}(y,T),\quad\boldsymbol{\psi^{\ast}}(y,x)=\boldsymbol{b}(y)\,F_{0}(x,T).
\end{gather*}
To determine the relationship between the vector fields $\boldsymbol{a}(x)$,
$\boldsymbol{b}(y)$ and TEs $F_{0}(x,T)$, $F_{0}^{\ast}(y,T)$,
recall that if the combined system $\mathrm{S+S_{ad}}$
is isolated, then $\dot{H}(x)=-\dot{h}(y)$; and if $T\neq0$, then $\dot{H}(x)+\dot{h}(y)\sim0$. 
Straightforward calculations show that
\[
\boldsymbol{a}(x)=\boldsymbol{\varphi}(x),\quad\boldsymbol{b}(y)=\boldsymbol{Q}(y),
\]
provided that $\boldsymbol{b}(y)\exp[-\beta h(y)]\underset{}{\rightarrow}\boldsymbol{0}$
as $\left|y\right|\rightarrow\infty$ and $\boldsymbol{a}(x)\exp[-\beta H(x)]\rightarrow\boldsymbol{0}$
as $\left|x\right|\rightarrow\infty$. As a result, we have the equations of motion 
\begin{gather}
\dot{x}=J_{x}\mathbb{\boldsymbol{\nabla}}_{x}H(x)+F_{0}^{\ast}(y,T)\boldsymbol{\boldsymbol{\varphi}}(x),\label{eq:GenNH}\\
\dot{y}=\boldsymbol{J}_{y}\boldsymbol{\nabla}_{y}h(y)-F_{0}(x,T)\boldsymbol{Q}(y),\nonumber 
\end{gather}
which are generalized NH equations. 

The Liouville equation associated
with the system \eqref{eq:GenNH} has the form $\partial_{t}\rho=-\mathcal{L}^{*}\rho$,
where $\mathcal{L}^{*}\rho=\boldsymbol{\nabla}_{x}\cdot(\dot{x}\rho)+\boldsymbol{\nabla}_{y}\cdot(\dot{y}\rho)$.
Invariant probability densities are determined
by the equation $\mathcal{L}^{*}\rho=0$. 
We claim that \emph{if $\boldsymbol{Q}(y)$ and $\boldsymbol{\varphi}(x)$
are the defined above vector fields, then the canonical density $\rho_{\infty}\propto\exp\left[-\beta H\left(x\right)\right]\cdot\exp\left[-\beta h\left(y\right)\right]$
is invariant for dynamics \eqref{eq:GenNH}, that is $\mathcal{L}^{*}\rho_{\infty}=0$.}
The proof is by direct calculation.

As a particular case, let $\boldsymbol{Q}(y)$ be an incompressible
vector field (i.e. $\boldsymbol{\nabla}_{y}\cdot\boldsymbol{Q}(y)=0$
for all $y\in\mathcal{M}_{\mathrm{ad}}$). Then we arrive at the NH
equations 
\begin{gather}
\dot{x}=J_{x}\mathbb{\boldsymbol{\nabla}}_{x}H(x)+\left(\boldsymbol{Q}(y)\cdot\boldsymbol{\nabla}_{y}h(y)\right)\boldsymbol{\boldsymbol{\varphi}}(x),\label{eq:NH}\\
\dot{y}=\boldsymbol{J}_{y}\boldsymbol{\nabla}_{y}h(y)-F_{0}(x,T)\boldsymbol{Q}(y).\nonumber 
\end{gather}

Now we include into our consideration the effect of the
thermal bath $\Sigma$ on $\mathrm{S_{ad}}$ dynamics, that is the
relationship $\boldsymbol{\nabla}_{y}h(y)\cdot\boldsymbol{\psi}^{\ast}=\lambda F_{0}^{\ast}(y,T)$.
Following the arguments and notations used to derive SDE \eqref{eq:deformed-StochEqM},
we arrive at the stochastic dynamics: 
\begin{gather}
\dot{x}=\boldsymbol{J}_{x}\mathbb{\boldsymbol{\nabla}}_{x}H(x)+F_{0}^{\ast}(y,T)\,\boldsymbol{\varphi}(x),\nonumber \\
\dot{y}=\boldsymbol{J}_{y}\boldsymbol{\nabla}_{y}h(y)-F_{0}(x,T)\,\boldsymbol{Q}(y)-\lambda\boldsymbol{\eta}(y)\circ\boldsymbol{\nabla}_{y}h(y)\nonumber \\
+\boldsymbol{\zeta}(y)\circ\boldsymbol{\xi}(t),\label{eq:GenNHL}
\end{gather}
which are generalized NHL equations\cite{SamoletovDettmannChaplain2007,LeimkuhlerNoorizadehTheil2009}.
In the particular case of an incompressible vector field $\boldsymbol{Q}(y)$
we get the NHL equations: 
\begin{gather}
\dot{x}=\boldsymbol{J}_{x}\mathbb{\boldsymbol{\nabla}}_{x}H(x)+\left(\boldsymbol{Q}(y)\cdot\boldsymbol{\nabla}_{y}h(y)\right)\,\boldsymbol{\varphi}(x),\nonumber \\
\dot{y}=\boldsymbol{J}_{y}\boldsymbol{\nabla}_{y}h(y)-F_{0}(x,T)\,\boldsymbol{Q}(y)-\lambda\boldsymbol{\eta}(y)\circ\boldsymbol{\nabla}_{y}h(y)\nonumber \\
+\boldsymbol{\zeta}(y)\circ\boldsymbol{\xi}(t),\label{eq:GenNHL-1}
\end{gather}

FPE corresponding to \eqref{eq:GenNHL} has the form $\partial_{t}\rho=\mathcal{F}^{*}\rho$,
where 
\begin{gather*}
\boldsymbol{\mathcal{F}}^{*}\rho=-\boldsymbol{J}_{x}\boldsymbol{\nabla}_{x}H(x)\cdot\boldsymbol{\boldsymbol{\nabla}}_{x}\rho-\boldsymbol{J}_{y}\boldsymbol{\nabla}_{y}h(y)\cdot\boldsymbol{\nabla}_{y}\rho\\
-F_{0}^{\ast}(y,T)\boldsymbol{\nabla}_{x}\cdot\left[\boldsymbol{\varphi}(x)\rho\right]+F_{0}(x,T)\boldsymbol{\nabla}_{y}\cdot\left[\boldsymbol{Q}(y)\rho\right]\\
+\lambda k_{B}T\boldsymbol{\nabla}_{y}\cdot[\boldsymbol{\eta}(y)\circ\boldsymbol{\nabla}_{y}\rho]+\lambda\boldsymbol{\boldsymbol{\nabla}}_{y}\cdot[\boldsymbol{\eta}(y)\circ\boldsymbol{\nabla}_{y}h(y)\rho].
\end{gather*}
Invariant probability density for the SDE \eqref{eq:GenNHL} is determined
by the equation $\mathcal{F}^{*}\rho=0$. 

We claim that \emph{if $\boldsymbol{Q}(y)$, $\boldsymbol{\varphi}(x)$,
and $\boldsymbol{\zeta}(y)$ are the defined above vector fields ,
then the canonical density, $\rho_{\infty}\propto\exp\left[-\beta H\left(x\right)\right]\cdot\exp\left[-\beta h\left(y\right)\right]$,
is invariant for the NHL dynamics \eqref{eq:GenNHL}, that is $\mathcal{F}^{*}\rho_{\infty}=0$.}
The proof is by direct calculation. 

Besides, we expect that this dynamics is ergodic \cite{mattingly2002ergodicity,LeimkuhlerNoorizadehTheil2009}.

Commonly used NH \cite{nose1984molecular,hoover1985canonical} and NHL \cite{SamoletovDettmannChaplain2007,LeimkuhlerNoorizadehTheil2009,SamoletovDettmannChaplain2010} thermostats are particular cases of thermostats given by \eqref{eq:NH}
and \eqref{eq:GenNHL-1} correspondingly. For example, by substituting
$\nicefrac{\zeta^{2}}{2Q}$ for $h(y),\,y=(\zeta,\eta)\in\mathbb{R}^{2}$,
$(-Q,0)$ for $\boldsymbol{Q}(y)$ and $(\mathbf{p},0)$ for $\boldsymbol{\varphi}(x)$
in \eqref{eq:GenNH} we get classical NH equations \cite{nose1984molecular,hoover1985canonical}.

It is worth to note that the case of the general TE can be considered
straightforwardly following the method of dynamic principle, as developed
above. Assume that
\begin{align*}
\boldsymbol{\nabla}_{x}H(x)\cdot\boldsymbol{\psi}(x,y) & =\sum_{l=0}^{L}g_{l}(x)F_{l}^{\ast}(y,T)\,(k_{B}T)^{2l},\\
\boldsymbol{\nabla}_{y}h(y)\cdot\boldsymbol{\psi}^{\ast}(y,x) & =-\sum_{l=0}^{L}g_{l}^{\ast}(y)F_{l}(x,T)\,(k_{B}T)^{2l}.
\end{align*}
These relationships must be valid for any $H(x)$ and $h(y)$. To specify
$\boldsymbol{\psi}(x,y)$ and $\boldsymbol{\psi}^{*}(y,x)$, we set
$g_{l}(x)=\boldsymbol{\boldsymbol{a}}_{l}(x)\cdot\boldsymbol{\nabla}_{x}H(x)$,
$g_{l}^{\ast}(y)=\boldsymbol{b}_{l}(y)\cdot\boldsymbol{\nabla}_{y}h(y)$,
from what follows that
$\boldsymbol{a}_{l}(x)=\boldsymbol{\boldsymbol{\varphi}}_{l}(x)$
and $\boldsymbol{b}_{l}(y)=\boldsymbol{Q}_{l}(y)$. Thus,
\begin{gather*}
\boldsymbol{\psi}(x,y)=\sum_{l=0}^{L}F_{l}^{\ast}(y,T)\,(k_{B}T)^{2l}\boldsymbol{\boldsymbol{\varphi}}_{l}(x),\\
\boldsymbol{\psi}^{\ast}(y,x)=-\sum_{l=0}^{L}F_{l}(x,T)\,(k_{B}T)^{2l}\boldsymbol{Q}_{l}(y).
\end{gather*}
Finally, we arrive at the deterministic equations of motion (modified Hamiltonian dynamics),
\begin{align}
\dot{x}= & J_{x}\mathbb{\boldsymbol{\nabla}}_{x}H(x)+\sum_{l=0}^{L}F_{l}^{\ast}(y,T)\,(k_{B}T)^{2l}\boldsymbol{\boldsymbol{\varphi}}_{l}(x),\nonumber \\
\dot{y}= & \boldsymbol{J}_{y}\boldsymbol{\nabla}_{y}h(y)-\sum_{l=0}^{L}F_{l}(x,T)\,(k_{B}T)^{2l}\boldsymbol{Q}_{l}(y).\label{eq:GenGenNH}
\end{align}
We will not discuss the equations \eqref{eq:GenGenNH} in detail and only note that the canonical measure is invariant for this dynamics, and a generalization to stochastic NHL type dynamics can be obtained. Strictly speaking, such a generalization is important since it simulates an equilibrium reservoir of the energy and ensures the ergodicity of dynamics. To outline a connection
between equations of motion \eqref{eq:GenGenNH} and known deterministic thermostats\cite{HooverHolian1996,HooverSprottPatra2015,Hoover2016,PatraBhattacharya2014,patra2015ergodic},
we provide the following simple example. Let $L=1$, $H(x)=\nicefrac{p^{2}}{2m}+V(q)$,
$h(y)=\nicefrac{\eta_{0}^{2}}{2Q_{0}}+\nicefrac{\eta_{1}^{2}}{2Q_{1}}$,
$\boldsymbol{\varphi}_{0}(x)=(p,0)$, $\boldsymbol{\varphi}_{1}(x)=(p^{3},0)$,
$\boldsymbol{Q}_{0}(y)=(-Q_{0},0,0,0)$, and $\boldsymbol{Q}_{1}(y)=(0,-Q_{1},0,0)$,
then
\begin{gather*}
\dot{p}=-V'(q)-\eta_{0}p-\eta_{1}k_{B}T\,p^{3},\\
\dot{q}=\frac{p}{m},\\
\dot{\eta}_{0}=Q_{0}\left(\frac{p^{2}}{m}-k_{B}T\right),\\
\dot{\eta}_{1}=Q_{1}\left(\frac{p^{4}}{m}-3k_{B}Tp^{2}\right)(k_{B}T)^{2},
\end{gather*}
the  dynamic equations equipped with the control of first two moments of the equilibrium kinetic energy\cite{HooverHolian1996,Hoover2016}. Similarly, we can obtain dynamic equations that control the configurational temperature moments. 

\section{Redesign of NHL thermostat}

In this Section we consider an alternative to the conventional NH and NHL thermostat schemes. 
This alternative (seen as a particular case of dynamical equations \eqref{eq:GenNHL}) 
is based on the consideration of physically reasonable chain of interactions, 
$\mathrm{S\leftrightsquigarrow S_{ad}}\leftrightsquigarrow\Sigma$, that is, the system $\mathrm{S_{ad}}$ is a buffer between the physical system $\mathrm{S}$ and the infinite energy reservoir $\mathrm{\Sigma}$. 

Consider the dynamical equations \eqref{eq:GenNH} and \eqref{eq:GenNHL},
and assume that $\boldsymbol{\nabla}_{x}\cdot\boldsymbol{\boldsymbol{\varphi}}(x)=0,\:\boldsymbol{\nabla}_{y}\cdot\boldsymbol{Q}(y)\neq0$.
Note, that these assumptions are opposite to the requirements for the NH and NHL dynamics, where $\boldsymbol{\nabla}_{x}\cdot\boldsymbol{\boldsymbol{\varphi}}(x)\neq0,\:\boldsymbol{\nabla}_{y}\cdot\boldsymbol{Q}(y)=0$.
We get

\begin{alignat}{1}
\dot{x} & =\boldsymbol{J}_{x}\mathbb{\boldsymbol{\nabla}}_{x}H(x)\nonumber \\
 & +\left[\boldsymbol{Q}(y)\cdot\boldsymbol{\nabla}_{y}h(y)-k_{B}T\,\boldsymbol{\nabla}_{y}\cdot\boldsymbol{Q}(y)\right]\,\boldsymbol{\varphi}(x),\nonumber \\
\dot{y} & =\boldsymbol{J}_{y}\boldsymbol{\nabla}_{y}h(y)-\left(\boldsymbol{\boldsymbol{\varphi}}(x)\cdot\boldsymbol{\nabla}_{x}H(x)\right)\boldsymbol{Q}(y),\label{eq:GenNH-Redesi}
\end{alignat}
and
\begin{flalign}
\dot{x} & =\boldsymbol{J}_{x}\mathbb{\boldsymbol{\nabla}}_{x}H(x)\nonumber \\
 & +\left[\boldsymbol{Q}(y)\cdot\boldsymbol{\nabla}_{y}h(y)-k_{B}T\,\boldsymbol{\nabla}_{y}\cdot\boldsymbol{Q}(y)\right]\,\boldsymbol{\varphi}(x),\nonumber \\
\dot{y} & =\boldsymbol{J}_{y}\boldsymbol{\nabla}_{y}h(y)-\left(\boldsymbol{\boldsymbol{\varphi}}(x)\cdot\boldsymbol{\nabla}_{x}H(x)\right)\,\boldsymbol{Q}(y)\nonumber \\
 & -\lambda\boldsymbol{\eta}(y)\circ\boldsymbol{\nabla}_{y}h(y)+\boldsymbol{\zeta}(y)\circ\boldsymbol{\xi}(t),\label{eq:GenNHL-Redesi}
\end{flalign}
where vector fields involved are such as indicated above. Thus, there is plenty of freedom in specification of particular thermostat equations of motion. 

To illustrate the redesigned NH and NHL thermostat dynamical systems (described by the equations \eqref{eq:GenNH-Redesi} and \eqref{eq:GenNHL-Redesi} correspondingly)
let us consider system $\mathrm{S}$ with the Hamiltonian function
$H\left(p,q\right)$,
\begin{gather*}
H\left(p,q\right)=\frac{p^{2}}{2m}+\frac{1}{2}m\omega^{2}q^{2},\quad x=(p,q)\in\mathbb{R}\times\mathbb{R},
\end{gather*}
that is a harmonic oscillator of mass $m$ and frequency $\omega$,
and system $\mathrm{S_{ad}}$ with the Hamiltonian function $h(v,u)$,
\begin{gather*}
h(v,u)=\frac{v^{2}}{2\mu},\quad y=(v,u)\in\mathbb{R}\times\mathbb{R},
\end{gather*}
that is a free particle of mass $\mu$. Harmonic oscillators are among
central instruments in analysis of many physical problems, classical
as well as quantum mechanical. It is known that generating the canonical
statistics for a harmonic oscillator is a hard problem. For example,
the NH scheme is proven to be non-ergodic\cite{legoll2007non} and the NHL scheme\cite{SamoletovDettmannChaplain2007,LeimkuhlerNoorizadehTheil2009}, and earlier the NHC scheme\cite{martyna1992nose},
was proposed to overcome this difficulty. Anyway, it is important for any dynamic thermostat
to correctly generate the canonical statistics for a harmonic oscillator.

The deterministic thermostat dynamics \eqref{eq:GenNH-Redesi} as
well as stochastic dynamics \eqref{eq:GenNHL-Redesi} allow a plethora
of further specifications. To be as close as possible to redesign
of original NH dynamics\cite{hoover1985canonical}, we set $\boldsymbol{Q}(y)=(v,0)$,
$\boldsymbol{\nabla}\cdot\boldsymbol{Q}=1$, and $\boldsymbol{\varphi}(x)=(\gamma,0)$,
where $\gamma$ is a dimensional parameter, $\boldsymbol{\nabla}\cdot\boldsymbol{\boldsymbol{\varphi}}=0$.
Thus, we arrive at the following equations of motion: 
\begin{alignat}{1}
\dot{p} & =-m\omega^{2}q+\gamma\left[\frac{v^{2}}{\mu}-k_{B}T\right],\nonumber \\
\dot{q} & =\frac{1}{m}p,\nonumber \\
\dot{v} & =-\gamma\frac{p}{m}v,\nonumber \\
\dot{u} & =\frac{v}{\mu};\label{eq:RNH}
\end{alignat}
and
\begin{alignat}{1}
\dot{p} & =-m\omega^{2}q+\gamma\left[\frac{v^{2}}{\mu}-k_{B}T\right],\nonumber \\
\dot{q} & =\frac{1}{m}p,\nonumber \\
\dot{v} & =-\gamma\frac{p}{m}v-\lambda\frac{v}{\mu}+\xi(t),\nonumber \\
\dot{u} & =\frac{v}{\mu};\label{eq:RNHL}
\end{alignat}
where $\boldsymbol{\zeta}=(1,0)$ and $\left\langle \xi(t)\xi(t')\right\rangle =2\lambda k_{\mathrm{B}}T\delta(t-t')$. Note, that  equations (\ref{eq:RNH}) and (\ref{eq:RNHL}) are redesign
of NH (denote RNH) and NHL (RNHL) thermostats correspondingly.

\begin{figure}[ht]
\includegraphics[width=5cm]{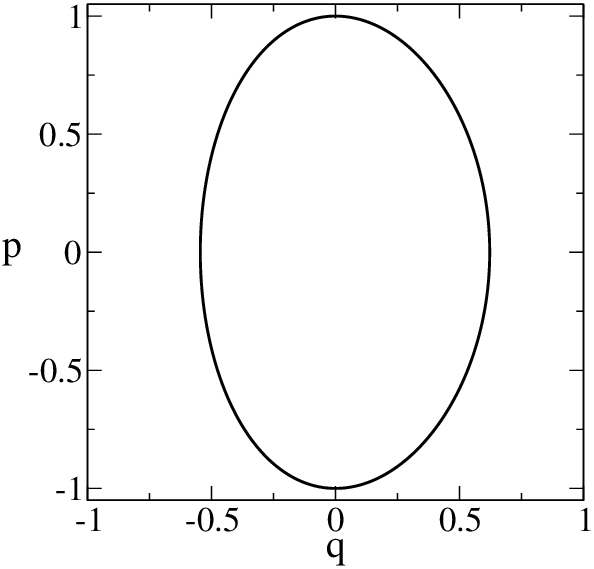}\caption{The Poincar{\'e} section (p,q) for deterministic dynamic thermostat \eqref{eq:RNH} where all system parameters are set equal to unity and initial condition are: $p=1,q=0,v=1$.\label{fig: 1}}
\end{figure}

System \eqref{eq:RNH} has two integrals of motion, that is
\begin{gather*}
I_{1}=v\exp\left(\gamma q\right)=\mathrm{const},\\
I_{2}=\frac{p^{2}}{2m}+\frac{1}{2}m\omega^{2}q^{2}+\frac{v^{2}}{2\mu}+\frac{\gamma k_{B}T}{m}q=\mathrm{const},
\end{gather*}
indicating the lack of ergodicity. For example, if all parameters of the system
\eqref{eq:RNH} are set equal to unity, $m=1,\omega=1,\mu=1,\gamma=1,k_{B}T=1$,
and initial conditions are $p=1,q=0,v=1$, then the phase trajectory is represented by the closed curve
and the Poincar{\'e} section (p,q) shown on Figure \ref{fig: 1}. This is expected from the existence
of two integrals of motion, that is $I_{1}$ and $I_{2}$. It is clear
that the trajectory does not explore the phase space available for
the harmonic oscillator. This ergodicity problem is not surprising,
the convenient NH dynamic suffer from the same problem. It is questionable  that the situation can be improved with a more complex $\boldsymbol{\varphi}$ and $\boldsymbol{Q}$,
for example,
$\boldsymbol{\varphi}=(\gamma_{1},\gamma_{2})$, $\boldsymbol{\varphi}=\left(\gamma_{1}m\omega^{2}q,\gamma_{2}\tfrac{1}{m}p\right)$,
$\boldsymbol{Q}=(v,u)$, and so on. If $\boldsymbol{\varphi}=(\gamma_{1},\gamma_{2})$,
then we get
\begin{alignat*}{1}
\dot{p} & =-m\omega^{2}q+\gamma_{1}\left[\frac{v^{2}}{\mu}-k_{B}T\right],\\
\dot{q} & =\frac{p}{m}+\gamma_{2}\left[\frac{v^{2}}{\mu}-k_{B}T\right],\\
\dot{v} & =-\left(\gamma_{1}\frac{p}{m}+\gamma_{2}\mu\omega^{2}q\right)v,\\
\dot{u} & =\frac{v}{\mu},
\end{alignat*}
and it is easy to show that this dynamics is not ergodic.

Our next illustration will be devoted to the system described by thermostat dynamical equations \eqref{eq:RNHL}. We will show, by means of numerical simulations, that a certain realization of the whole length chain of physically
reasonable interactions, that is $\mathrm{S\leftrightsquigarrow S_{ad}}\leftrightsquigarrow\Sigma$,
generates the correct statistics.

Let us consider the case when all parameters of the system \eqref{eq:RNHL} are set equal to unity, $m=1,\omega=1,\mu=1,\gamma=1,k_{B}T=1,\lambda=1$,
and the initial conditions are: $p=0,q=0,v=0$. Phase trajectories of length $10^6$ are generated using the Euler method with a time step of $\Delta t=0.0005$. We have repeated simulations using the fourth-order Runge-Kutta method with a random contribution held once for the entire interval from $t$ to $ t+\Delta t$, and arrive at the same result.

\begin{figure}[ht]
	\begin{tabular}{c}
		\rule[-1ex]{0pt}{2.5ex} \includegraphics[width=5cm]{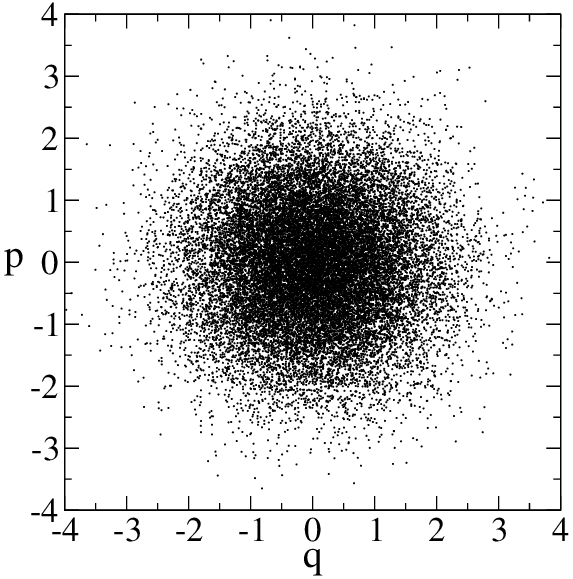}(\textbf{\large{}a})\bigskip \bigskip \\ 
		\rule[-1ex]{0pt}{2.5ex} \includegraphics[width=5cm,height=4cm]{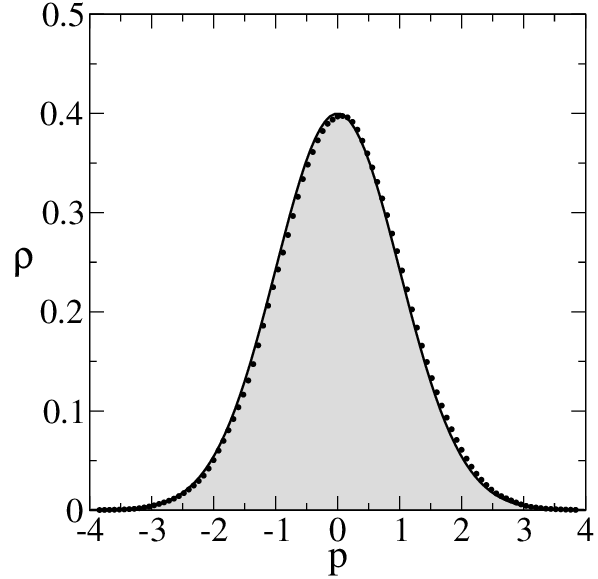}(\textbf{\large{}b}) \bigskip \bigskip \\ 
		\rule[-1ex]{0pt}{2.5ex}  \includegraphics[width=5cm,height=4cm]{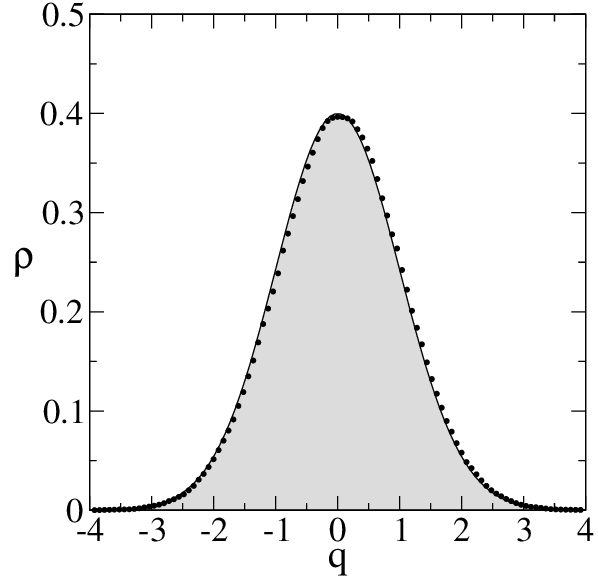}(\textbf{\large{}c})\\ 
	\end{tabular} 
\caption{(a) The Poincar\'e section (p,q) for a harmonic oscillator coupled
to the redesigned NHL thermostat \eqref{eq:RNHL}. (b) The generated
momentum density function (dots) as compared to the analytical result (solid cover filled in gray). (c) The same for the generated position density function. \label{fig:2}}
\end{figure}

Figure \ref{fig:2}(a) shows the Poincar\'e section (p,q) for a harmonic oscillator equipped with the temperature control tool \eqref{eq:RNHL}. This figure demonstrates that the trajectory generates proper sampling of the full phase space of the harmonic oscillator. Figures \ref{fig:2}(b) and \ref{fig:2}(c) show the momentum and position distribution functions from simulations as 
compared with the exact analytical expressions. In both cases, the
Gaussian distribution is generated in agreement with the theoretical prediction. Presented results serve as an evidence of ergodic sampling the canonical statistics. 

\begin{figure}[]
\includegraphics[width=5cm,height=4cm]{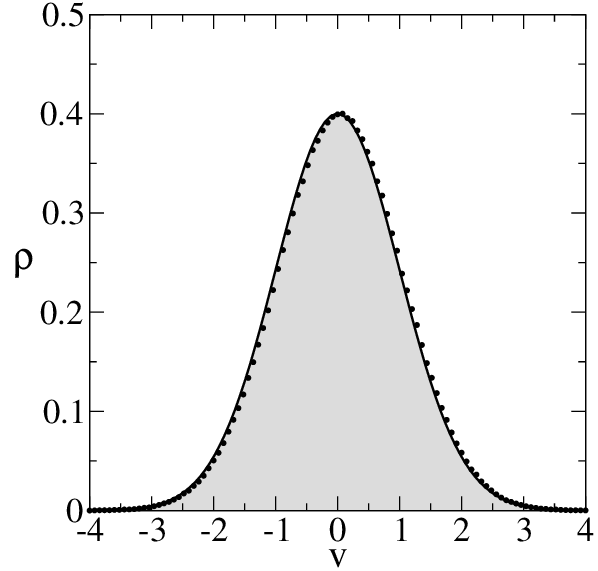}\caption{The density function for momentum $v$ (in the system $\mathrm{S_{ad}}$) from simulations (dots) as compared with the analytical solution (solid cover filled in gray).
\label{fig:3}}

\end{figure}

A key difference between the NHL and RNHL schemes is that the latter relates
the temperature control tool to the system $\mathrm{S_{ad}}$ rather than to the system $\mathrm{S}$, and the corresponding variable, $v$, must be Gaussian, according to the equations \eqref{eq:RNHL}. Thus, it is important that the RNHL dynamical equations properly generate the Gaussian
statistics of $v$ variable. Figure \ref{fig:3} shows the $v$-distribution function from simulations as compared with the exact analytical solution and indicates a good agreement between them. 

\section{Conclusion}

In conclusion, we emphasize that the method proposed in this work
is based on the fundamental laws of statistical physics and offers
a unified approach in developing stochastic and deterministic thermostats. For clarity of presentation we have illustrated our method using a few simple TEs and restricted our consideration by Markov dynamics. The presented method allowed us to obtain a wide spectrum of stochastic and deterministic dynamical systems with the invariant canonical measure. We note that the idea
of presented method is general and adaptable to a variety of TEs so
that it can be used to produce thermostats of novel types. For example the thermostat for the system with non-Markov dynamics, i.e. the one described by the equation $\boldsymbol{\nabla}H(x(t))\cdot\dot{x}(t)\propto\int_{0}^{t}dt'G(t-t')F(x(t'),T)$. As a second example of new type of thermostats we can mention the one for the gradient dynamical system. 

We realize that non-trivial new thermostats should be verified by
test simulations. In our follow up work we will focus on these and other applications of the presented method. 
\begin{acknowledgments}
This work has been supported by the BBSRC grant BB/K002430/1 to BV. 
\end{acknowledgments}

\bibliographystyle{apsrev4-1}
\bibliography{thermo}

\begin{thebibliography}{53}%
\makeatletter
\providecommand \@ifxundefined [1]{%
 \@ifx{#1\undefined}
}%
\providecommand \@ifnum [1]{%
 \ifnum #1\expandafter \@firstoftwo
 \else \expandafter \@secondoftwo
 \fi
}%
\providecommand \@ifx [1]{%
 \ifx #1\expandafter \@firstoftwo
 \else \expandafter \@secondoftwo
 \fi
}%
\providecommand \natexlab [1]{#1}%
\providecommand \enquote  [1]{``#1''}%
\providecommand \bibnamefont  [1]{#1}%
\providecommand \bibfnamefont [1]{#1}%
\providecommand \citenamefont [1]{#1}%
\providecommand \href@noop [0]{\@secondoftwo}%
\providecommand \href [0]{\begingroup \@sanitize@url \@href}%
\providecommand \@href[1]{\@@startlink{#1}\@@href}%
\providecommand \@@href[1]{\endgroup#1\@@endlink}%
\providecommand \@sanitize@url [0]{\catcode `\\12\catcode `\$12\catcode
  `\&12\catcode `\#12\catcode `\^12\catcode `\_12\catcode `\%12\relax}%
\providecommand \@@startlink[1]{}%
\providecommand \@@endlink[0]{}%
\providecommand \url  [0]{\begingroup\@sanitize@url \@url }%
\providecommand \@url [1]{\endgroup\@href {#1}{\urlprefix }}%
\providecommand \urlprefix  [0]{URL }%
\providecommand \Eprint [0]{\href }%
\providecommand \doibase [0]{http://dx.doi.org/}%
\providecommand \selectlanguage [0]{\@gobble}%
\providecommand \bibinfo  [0]{\@secondoftwo}%
\providecommand \bibfield  [0]{\@secondoftwo}%
\providecommand \translation [1]{[#1]}%
\providecommand \BibitemOpen [0]{}%
\providecommand \bibitemStop [0]{}%
\providecommand \bibitemNoStop [0]{.\EOS\space}%
\providecommand \EOS [0]{\spacefactor3000\relax}%
\providecommand \BibitemShut  [1]{\csname bibitem#1\endcsname}%
\let\auto@bib@innerbib\@empty
\bibitem [{\citenamefont {Allen}\ and\ \citenamefont
  {Tildesley}(1989)}]{AllenTildesley1989}%
  \BibitemOpen
  \bibfield  {author} {\bibinfo {author} {\bibfnamefont {M.~P.}\ \bibnamefont
  {Allen}}\ and\ \bibinfo {author} {\bibfnamefont {D.~J.}\ \bibnamefont
  {Tildesley}},\ }\href@noop {} {\emph {\bibinfo {title} {Computer simulation
  of liquids}}}\ (\bibinfo  {publisher} {Oxford University Press},\ \bibinfo
  {year} {1989})\BibitemShut {NoStop}%
\bibitem [{\citenamefont {Frenkel}\ and\ \citenamefont
  {Smit}(2002)}]{frenkel2002understanding}%
  \BibitemOpen
  \bibfield  {author} {\bibinfo {author} {\bibfnamefont {D.}~\bibnamefont
  {Frenkel}}\ and\ \bibinfo {author} {\bibfnamefont {B.}~\bibnamefont {Smit}},\
  }\href@noop {} {\emph {\bibinfo {title} {Understanding molecular simulation:
  from algorithms to applications}}}\ (\bibinfo  {publisher} {Elsevier},\
  \bibinfo {year} {2002})\BibitemShut {NoStop}%
\bibitem [{\citenamefont {Leimkuhler}\ and\ \citenamefont
  {Matthews}(2015)}]{LeimkuhlerMatthews2015}%
  \BibitemOpen
  \bibfield  {author} {\bibinfo {author} {\bibfnamefont {B.}~\bibnamefont
  {Leimkuhler}}\ and\ \bibinfo {author} {\bibfnamefont {C.}~\bibnamefont
  {Matthews}},\ }\href@noop {} {\emph {\bibinfo {title} {Molecular Dynamics:
  with deterministic and stochastic numerical methods}}}\ (\bibinfo
  {publisher} {Springer},\ \bibinfo {year} {2015})\BibitemShut {NoStop}%
\bibitem [{\citenamefont {Tuckerman}(2010)}]{tuckerman2010statistical}%
  \BibitemOpen
  \bibfield  {author} {\bibinfo {author} {\bibfnamefont {M.}~\bibnamefont
  {Tuckerman}},\ }\href@noop {} {\emph {\bibinfo {title} {Statistical
  mechanics: theory and molecular simulation}}}\ (\bibinfo  {publisher} {Oxford
  University Press},\ \bibinfo {year} {2010})\BibitemShut {NoStop}%
\bibitem [{\citenamefont {{Jepps}}\ and\ \citenamefont
  {{Rondoni}}(2010)}]{JeppsRondoni2010}%
  \BibitemOpen
  \bibfield  {author} {\bibinfo {author} {\bibfnamefont {O.~G.}\ \bibnamefont
  {{Jepps}}}\ and\ \bibinfo {author} {\bibfnamefont {L.}~\bibnamefont
  {{Rondoni}}},\ }\href {\doibase 10.1088/1751-8113/43/13/133001} {\bibfield
  {journal} {\bibinfo  {journal} {J. Phys. A: Math. Gen.}\ }\textbf {\bibinfo
  {volume} {43}},\ \bibinfo {pages} {133001} (\bibinfo {year}
  {2010})}\BibitemShut {NoStop}%
\bibitem [{\citenamefont {{Bussi}}\ \emph {et~al.}(2007)\citenamefont
  {{Bussi}}, \citenamefont {{Donadio}},\ and\ \citenamefont
  {{Parrinello}}}]{BussiDonadioParrinello2007}%
  \BibitemOpen
  \bibfield  {author} {\bibinfo {author} {\bibfnamefont {G.}~\bibnamefont
  {{Bussi}}}, \bibinfo {author} {\bibfnamefont {D.}~\bibnamefont {{Donadio}}},
  \ and\ \bibinfo {author} {\bibfnamefont {M.}~\bibnamefont {{Parrinello}}},\
  }\href {\doibase 10.1063/1.2408420} {\bibfield  {journal} {\bibinfo
  {journal} {J. Chem. Phys.}\ }\textbf {\bibinfo {volume} {126}},\ \bibinfo
  {pages} {014101} (\bibinfo {year} {2007})}\BibitemShut {NoStop}%
\bibitem [{\citenamefont {Samoletov}\ \emph {et~al.}(2007)\citenamefont
  {Samoletov}, \citenamefont {Dettmann},\ and\ \citenamefont
  {Chaplain}}]{SamoletovDettmannChaplain2007}%
  \BibitemOpen
  \bibfield  {author} {\bibinfo {author} {\bibfnamefont {A.}~\bibnamefont
  {Samoletov}}, \bibinfo {author} {\bibfnamefont {C.}~\bibnamefont {Dettmann}},
  \ and\ \bibinfo {author} {\bibfnamefont {M.}~\bibnamefont {Chaplain}},\
  }\href {\doibase 10.1007/s10955-007-9365-2} {\bibfield  {journal} {\bibinfo
  {journal} {J. Stat. Phys.}\ }\textbf {\bibinfo {volume} {128}},\ \bibinfo
  {pages} {1321} (\bibinfo {year} {2007})}\BibitemShut {NoStop}%
\bibitem [{\citenamefont {{Leimkuhler}}\ \emph {et~al.}(2009)\citenamefont
  {{Leimkuhler}}, \citenamefont {{Noorizadeh}},\ and\ \citenamefont
  {{Theil}}}]{LeimkuhlerNoorizadehTheil2009}%
  \BibitemOpen
  \bibfield  {author} {\bibinfo {author} {\bibfnamefont {B.}~\bibnamefont
  {{Leimkuhler}}}, \bibinfo {author} {\bibfnamefont {E.}~\bibnamefont
  {{Noorizadeh}}}, \ and\ \bibinfo {author} {\bibfnamefont {F.}~\bibnamefont
  {{Theil}}},\ }\href {\doibase 10.1007/s10955-009-9734-0} {\bibfield
  {journal} {\bibinfo  {journal} {J. Stat. Phys.}\ }\textbf {\bibinfo {volume}
  {135}},\ \bibinfo {pages} {261} (\bibinfo {year} {2009})}\BibitemShut
  {NoStop}%
\bibitem [{\citenamefont {Samoletov}\ \emph {et~al.}(2010)\citenamefont
  {Samoletov}, \citenamefont {Dettmann},\ and\ \citenamefont
  {Chaplain}}]{SamoletovDettmannChaplain2010}%
  \BibitemOpen
  \bibfield  {author} {\bibinfo {author} {\bibfnamefont {A.}~\bibnamefont
  {Samoletov}}, \bibinfo {author} {\bibfnamefont {C.}~\bibnamefont {Dettmann}},
  \ and\ \bibinfo {author} {\bibfnamefont {M.}~\bibnamefont {Chaplain}},\
  }\href {\doibase 10.1063/1.3453655} {\bibfield  {journal} {\bibinfo
  {journal} {J. Chem. Phys.}\ }\textbf {\bibinfo {volume} {132}},\ \bibinfo
  {pages} {246101} (\bibinfo {year} {2010})}\BibitemShut {NoStop}%
\bibitem [{\citenamefont {Leimkuhler}(2010)}]{Leimkuhler2010}%
  \BibitemOpen
  \bibfield  {author} {\bibinfo {author} {\bibfnamefont {B.}~\bibnamefont
  {Leimkuhler}},\ }\href {\doibase 10.1103/PhysRevE.81.026703} {\bibfield
  {journal} {\bibinfo  {journal} {Phys. Rev. E}\ }\textbf {\bibinfo {volume}
  {81}},\ \bibinfo {pages} {026703} (\bibinfo {year} {2010})}\BibitemShut
  {NoStop}%
\bibitem [{\citenamefont {Bajars}\ \emph {et~al.}(2011)\citenamefont {Bajars},
  \citenamefont {Frank},\ and\ \citenamefont
  {Leimkuhler}}]{BajarsFrankLeimkuhler2011}%
  \BibitemOpen
  \bibfield  {author} {\bibinfo {author} {\bibfnamefont {J.}~\bibnamefont
  {Bajars}}, \bibinfo {author} {\bibfnamefont {J.}~\bibnamefont {Frank}}, \
  and\ \bibinfo {author} {\bibfnamefont {B.}~\bibnamefont {Leimkuhler}},\
  }\href {\doibase 10.1140/epjst/e2011-01522-0} {\bibfield  {journal} {\bibinfo
   {journal} {Eur. Phys. J. Special Topics}\ }\textbf {\bibinfo {volume}
  {200}},\ \bibinfo {pages} {131} (\bibinfo {year} {2011})}\BibitemShut
  {NoStop}%
\bibitem [{\citenamefont {Di~Pierro}\ \emph {et~al.}(2015)\citenamefont
  {Di~Pierro}, \citenamefont {Elber},\ and\ \citenamefont
  {Leimkuhler}}]{DiPierroElberLeimkuhler2015}%
  \BibitemOpen
  \bibfield  {author} {\bibinfo {author} {\bibfnamefont {M.}~\bibnamefont
  {Di~Pierro}}, \bibinfo {author} {\bibfnamefont {R.}~\bibnamefont {Elber}}, \
  and\ \bibinfo {author} {\bibfnamefont {B.}~\bibnamefont {Leimkuhler}},\
  }\href {\doibase 10.1021/acs.jctc.5b00648} {\bibfield  {journal} {\bibinfo
  {journal} {J. Chem. Theory Comput.}\ }\textbf {\bibinfo {volume} {11}},\
  \bibinfo {pages} {5624} (\bibinfo {year} {2015})}\BibitemShut {NoStop}%
\bibitem [{\citenamefont {Dittmar}\ and\ \citenamefont
  {Kusalik}(2014)}]{Dittmar2014}%
  \BibitemOpen
  \bibfield  {author} {\bibinfo {author} {\bibfnamefont {H.}~\bibnamefont
  {Dittmar}}\ and\ \bibinfo {author} {\bibfnamefont {P.~G.}\ \bibnamefont
  {Kusalik}},\ }\href {\doibase 10.1103/PhysRevLett.112.195701} {\bibfield
  {journal} {\bibinfo  {journal} {Phys. Rev. Lett.}\ }\textbf {\bibinfo
  {volume} {112}},\ \bibinfo {pages} {195701} (\bibinfo {year}
  {2014})}\BibitemShut {NoStop}%
\bibitem [{\citenamefont {Ness}\ \emph {et~al.}(2016)\citenamefont {Ness},
  \citenamefont {Genina}, \citenamefont {Stella}, \citenamefont {Lorenz},\ and\
  \citenamefont {Kantorovich}}]{ness2016nonequilibrium}%
  \BibitemOpen
  \bibfield  {author} {\bibinfo {author} {\bibfnamefont {H.}~\bibnamefont
  {Ness}}, \bibinfo {author} {\bibfnamefont {A.}~\bibnamefont {Genina}},
  \bibinfo {author} {\bibfnamefont {L.}~\bibnamefont {Stella}}, \bibinfo
  {author} {\bibfnamefont {C.}~\bibnamefont {Lorenz}}, \ and\ \bibinfo {author}
  {\bibfnamefont {L.}~\bibnamefont {Kantorovich}},\ }\href@noop {} {\bibfield
  {journal} {\bibinfo  {journal} {Phys. Rev. B}\ }\textbf {\bibinfo {volume}
  {93}},\ \bibinfo {pages} {174303} (\bibinfo {year} {2016})}\BibitemShut
  {NoStop}%
\bibitem [{\citenamefont {Dittmar}\ and\ \citenamefont
  {Kusalik}(2016)}]{dittmar2016ordering}%
  \BibitemOpen
  \bibfield  {author} {\bibinfo {author} {\bibfnamefont {H.~R.}\ \bibnamefont
  {Dittmar}}\ and\ \bibinfo {author} {\bibfnamefont {P.~G.}\ \bibnamefont
  {Kusalik}},\ }\href@noop {} {\bibfield  {journal} {\bibinfo  {journal} {J.
  Chem. Phys.}\ }\textbf {\bibinfo {volume} {145}},\ \bibinfo {pages} {134504}
  (\bibinfo {year} {2016})}\BibitemShut {NoStop}%
\bibitem [{\citenamefont {Stella}\ \emph {et~al.}(2014)\citenamefont {Stella},
  \citenamefont {Lorenz},\ and\ \citenamefont {Kantorovich}}]{Stella2014}%
  \BibitemOpen
  \bibfield  {author} {\bibinfo {author} {\bibfnamefont {L.}~\bibnamefont
  {Stella}}, \bibinfo {author} {\bibfnamefont {C.~D.}\ \bibnamefont {Lorenz}},
  \ and\ \bibinfo {author} {\bibfnamefont {L.}~\bibnamefont {Kantorovich}},\
  }\href {\doibase 10.1103/PhysRevB.89.134303} {\bibfield  {journal} {\bibinfo
  {journal} {Phys. Rev. B}\ }\textbf {\bibinfo {volume} {89}},\ \bibinfo
  {pages} {134303} (\bibinfo {year} {2014})}\BibitemShut {NoStop}%
\bibitem [{\citenamefont {{Lepri}}\ \emph {et~al.}(1997)\citenamefont
  {{Lepri}}, \citenamefont {{Livi}},\ and\ \citenamefont
  {{Politi}}}]{LepriLiviPoliti1997}%
  \BibitemOpen
  \bibfield  {author} {\bibinfo {author} {\bibfnamefont {S.}~\bibnamefont
  {{Lepri}}}, \bibinfo {author} {\bibfnamefont {R.}~\bibnamefont {{Livi}}}, \
  and\ \bibinfo {author} {\bibfnamefont {A.}~\bibnamefont {{Politi}}},\ }\href
  {\doibase 10.1103/PhysRevLett.78.1896} {\bibfield  {journal} {\bibinfo
  {journal} {Phys. Rev. Lett.}\ }\textbf {\bibinfo {volume} {78}},\ \bibinfo
  {pages} {1896} (\bibinfo {year} {1997})}\BibitemShut {NoStop}%
\bibitem [{\citenamefont {Pastorino}\ \emph {et~al.}(2007)\citenamefont
  {Pastorino}, \citenamefont {Kreer}, \citenamefont {M{\"u}ller},\ and\
  \citenamefont {Binder}}]{pastorino2007comparison}%
  \BibitemOpen
  \bibfield  {author} {\bibinfo {author} {\bibfnamefont {C.}~\bibnamefont
  {Pastorino}}, \bibinfo {author} {\bibfnamefont {T.}~\bibnamefont {Kreer}},
  \bibinfo {author} {\bibfnamefont {M.}~\bibnamefont {M{\"u}ller}}, \ and\
  \bibinfo {author} {\bibfnamefont {K.}~\bibnamefont {Binder}},\ }\href@noop {}
  {\bibfield  {journal} {\bibinfo  {journal} {Phys. Rev. E}\ }\textbf {\bibinfo
  {volume} {76}},\ \bibinfo {pages} {026706} (\bibinfo {year}
  {2007})}\BibitemShut {NoStop}%
\bibitem [{\citenamefont {Ciccotti}\ and\ \citenamefont
  {Ferrario}(2016)}]{ciccotti2016non}%
  \BibitemOpen
  \bibfield  {author} {\bibinfo {author} {\bibfnamefont {G.}~\bibnamefont
  {Ciccotti}}\ and\ \bibinfo {author} {\bibfnamefont {M.}~\bibnamefont
  {Ferrario}},\ }\href@noop {} {\bibfield  {journal} {\bibinfo  {journal} {Mol.
  Simul.}\ }\textbf {\bibinfo {volume} {42}},\ \bibinfo {pages} {1385}
  (\bibinfo {year} {2016})}\BibitemShut {NoStop}%
\bibitem [{\citenamefont {Bianca}(2012)}]{bianca2012thermostatted}%
  \BibitemOpen
  \bibfield  {author} {\bibinfo {author} {\bibfnamefont {C.}~\bibnamefont
  {Bianca}},\ }\href@noop {} {\bibfield  {journal} {\bibinfo  {journal} {Phys.
  Life Rev.}\ }\textbf {\bibinfo {volume} {9}},\ \bibinfo {pages} {359}
  (\bibinfo {year} {2012})}\BibitemShut {NoStop}%
\bibitem [{\citenamefont {Samoletov}\ and\ \citenamefont
  {Vasiev}(2013)}]{SamoletovVasiev2013}%
  \BibitemOpen
  \bibfield  {author} {\bibinfo {author} {\bibfnamefont {A.}~\bibnamefont
  {Samoletov}}\ and\ \bibinfo {author} {\bibfnamefont {B.}~\bibnamefont
  {Vasiev}},\ }\href {\doibase 10.1016/j.aml.2012.03.035} {\bibfield  {journal}
  {\bibinfo  {journal} {Appl. Math. Lett.}\ }\textbf {\bibinfo {volume} {26}},\
  \bibinfo {pages} {73} (\bibinfo {year} {2013})}\BibitemShut {NoStop}%
\bibitem [{\citenamefont {Chow}\ \emph {et~al.}(2005)\citenamefont {Chow},
  \citenamefont {Ram}, \citenamefont {Boker}, \citenamefont {Fujita},\ and\
  \citenamefont {Clore}}]{chow2005emotion}%
  \BibitemOpen
  \bibfield  {author} {\bibinfo {author} {\bibfnamefont {S.-M.}\ \bibnamefont
  {Chow}}, \bibinfo {author} {\bibfnamefont {N.}~\bibnamefont {Ram}}, \bibinfo
  {author} {\bibfnamefont {S.~M.}\ \bibnamefont {Boker}}, \bibinfo {author}
  {\bibfnamefont {F.}~\bibnamefont {Fujita}}, \ and\ \bibinfo {author}
  {\bibfnamefont {G.}~\bibnamefont {Clore}},\ }\href@noop {} {\bibfield
  {journal} {\bibinfo  {journal} {Emotion}\ }\textbf {\bibinfo {volume} {5}},\
  \bibinfo {pages} {208} (\bibinfo {year} {2005})}\BibitemShut {NoStop}%
\bibitem [{\citenamefont {Tang}\ \emph {et~al.}(2016)\citenamefont {Tang},
  \citenamefont {Li}, \citenamefont {Li}, \citenamefont {Deng},\ and\
  \citenamefont {Karniadakis}}]{tang2016non}%
  \BibitemOpen
  \bibfield  {author} {\bibinfo {author} {\bibfnamefont {Y.-H.}\ \bibnamefont
  {Tang}}, \bibinfo {author} {\bibfnamefont {Z.}~\bibnamefont {Li}}, \bibinfo
  {author} {\bibfnamefont {X.}~\bibnamefont {Li}}, \bibinfo {author}
  {\bibfnamefont {M.}~\bibnamefont {Deng}}, \ and\ \bibinfo {author}
  {\bibfnamefont {G.~E.}\ \bibnamefont {Karniadakis}},\ }\href@noop {}
  {\bibfield  {journal} {\bibinfo  {journal} {Macromolecules}\ }\textbf
  {\bibinfo {volume} {49}},\ \bibinfo {pages} {2895} (\bibinfo {year}
  {2016})}\BibitemShut {NoStop}%
\bibitem [{\citenamefont {Mones}\ \emph {et~al.}(2015)\citenamefont {Mones},
  \citenamefont {Jones}, \citenamefont {Goetz}, \citenamefont {Laino},
  \citenamefont {Walker}, \citenamefont {Leimkuhler}, \citenamefont {Csanyi},\
  and\ \citenamefont {Bernstein}}]{MonesJonesGoetzEtAl2015}%
  \BibitemOpen
  \bibfield  {author} {\bibinfo {author} {\bibfnamefont {L.}~\bibnamefont
  {Mones}}, \bibinfo {author} {\bibfnamefont {A.}~\bibnamefont {Jones}},
  \bibinfo {author} {\bibfnamefont {A.~W.}\ \bibnamefont {Goetz}}, \bibinfo
  {author} {\bibfnamefont {T.}~\bibnamefont {Laino}}, \bibinfo {author}
  {\bibfnamefont {R.~C.}\ \bibnamefont {Walker}}, \bibinfo {author}
  {\bibfnamefont {B.}~\bibnamefont {Leimkuhler}}, \bibinfo {author}
  {\bibfnamefont {G.}~\bibnamefont {Csanyi}}, \ and\ \bibinfo {author}
  {\bibfnamefont {N.}~\bibnamefont {Bernstein}},\ }\href {\doibase
  10.1002/jcc.23839} {\bibfield  {journal} {\bibinfo  {journal} {J. Comput.
  Chem.}\ }\textbf {\bibinfo {volume} {36}},\ \bibinfo {pages} {633} (\bibinfo
  {year} {2015})}\BibitemShut {NoStop}%
\bibitem [{\citenamefont {Fritz}\ \emph {et~al.}(2011)\citenamefont {Fritz},
  \citenamefont {Koschke}, \citenamefont {Harmandaris}, \citenamefont {van~der
  Vegt},\ and\ \citenamefont {Kremer}}]{fritz2011multiscale}%
  \BibitemOpen
  \bibfield  {author} {\bibinfo {author} {\bibfnamefont {D.}~\bibnamefont
  {Fritz}}, \bibinfo {author} {\bibfnamefont {K.}~\bibnamefont {Koschke}},
  \bibinfo {author} {\bibfnamefont {V.~A.}\ \bibnamefont {Harmandaris}},
  \bibinfo {author} {\bibfnamefont {N.~F.}\ \bibnamefont {van~der Vegt}}, \
  and\ \bibinfo {author} {\bibfnamefont {K.}~\bibnamefont {Kremer}},\
  }\href@noop {} {\bibfield  {journal} {\bibinfo  {journal} {Phys. Chem. Chem.
  Phys.}\ }\textbf {\bibinfo {volume} {13}},\ \bibinfo {pages} {10412}
  (\bibinfo {year} {2011})}\BibitemShut {NoStop}%
\bibitem [{\citenamefont {Praprotnik}\ \emph {et~al.}(2008)\citenamefont
  {Praprotnik}, \citenamefont {Site},\ and\ \citenamefont
  {Kremer}}]{praprotnik2008multiscale}%
  \BibitemOpen
  \bibfield  {author} {\bibinfo {author} {\bibfnamefont {M.}~\bibnamefont
  {Praprotnik}}, \bibinfo {author} {\bibfnamefont {L.~D.}\ \bibnamefont
  {Site}}, \ and\ \bibinfo {author} {\bibfnamefont {K.}~\bibnamefont
  {Kremer}},\ }\href@noop {} {\bibfield  {journal} {\bibinfo  {journal} {Annu.
  Rev. Phys. Chem.}\ }\textbf {\bibinfo {volume} {59}},\ \bibinfo {pages} {545}
  (\bibinfo {year} {2008})}\BibitemShut {NoStop}%
\bibitem [{\citenamefont {Chen}\ \emph {et~al.}(2016)\citenamefont {Chen},
  \citenamefont {Ding}, \citenamefont {Li}, \citenamefont {Zhang},\ and\
  \citenamefont {Carin}}]{Chen2016}%
  \BibitemOpen
  \bibfield  {author} {\bibinfo {author} {\bibfnamefont {C.}~\bibnamefont
  {Chen}}, \bibinfo {author} {\bibfnamefont {N.}~\bibnamefont {Ding}}, \bibinfo
  {author} {\bibfnamefont {C.}~\bibnamefont {Li}}, \bibinfo {author}
  {\bibfnamefont {Y.}~\bibnamefont {Zhang}}, \ and\ \bibinfo {author}
  {\bibfnamefont {L.}~\bibnamefont {Carin}},\ }in\ \href@noop {} {\emph
  {\bibinfo {booktitle} {Advances In Neural Information Processing Systems}}}\
  (\bibinfo {year} {2016})\ pp.\ \bibinfo {pages} {2937--2945}\BibitemShut
  {NoStop}%
\bibitem [{\citenamefont {Leimkuhler}\ and\ \citenamefont
  {Shang}(2016)}]{Leimkuhler2016}%
  \BibitemOpen
  \bibfield  {author} {\bibinfo {author} {\bibfnamefont {B.}~\bibnamefont
  {Leimkuhler}}\ and\ \bibinfo {author} {\bibfnamefont {X.}~\bibnamefont
  {Shang}},\ }\href@noop {} {\bibfield  {journal} {\bibinfo  {journal} {SIAM J.
  Sci. Comput.}\ }\textbf {\bibinfo {volume} {38}},\ \bibinfo {pages} {A712}
  (\bibinfo {year} {2016})}\BibitemShut {NoStop}%
\bibitem [{\citenamefont {Ding}\ \emph {et~al.}(2014)\citenamefont {Ding},
  \citenamefont {Fang}, \citenamefont {Babbush}, \citenamefont {Chen},
  \citenamefont {Skeel},\ and\ \citenamefont {Neven}}]{Ding2014}%
  \BibitemOpen
  \bibfield  {author} {\bibinfo {author} {\bibfnamefont {N.}~\bibnamefont
  {Ding}}, \bibinfo {author} {\bibfnamefont {Y.}~\bibnamefont {Fang}}, \bibinfo
  {author} {\bibfnamefont {R.}~\bibnamefont {Babbush}}, \bibinfo {author}
  {\bibfnamefont {C.}~\bibnamefont {Chen}}, \bibinfo {author} {\bibfnamefont
  {R.~D.}\ \bibnamefont {Skeel}}, \ and\ \bibinfo {author} {\bibfnamefont
  {H.}~\bibnamefont {Neven}},\ }in\ \href
  {http://papers.nips.cc/paper/5592-bayesian-sampling-using-stochastic-gradient-thermostats.pdf}
  {\emph {\bibinfo {booktitle} {Advances in Neural Information Processing
  Systems 27}}}\ (\bibinfo  {publisher} {Curran Associates, Inc.},\ \bibinfo
  {year} {2014})\ pp.\ \bibinfo {pages} {3203--3211}\BibitemShut {NoStop}%
\bibitem [{\citenamefont {Shang}\ \emph {et~al.}(2015)\citenamefont {Shang},
  \citenamefont {Zhu}, \citenamefont {Leimkuhler},\ and\ \citenamefont
  {Storkey}}]{NIPS2015_5978}%
  \BibitemOpen
  \bibfield  {author} {\bibinfo {author} {\bibfnamefont {X.}~\bibnamefont
  {Shang}}, \bibinfo {author} {\bibfnamefont {Z.}~\bibnamefont {Zhu}}, \bibinfo
  {author} {\bibfnamefont {B.}~\bibnamefont {Leimkuhler}}, \ and\ \bibinfo
  {author} {\bibfnamefont {A.~J.}\ \bibnamefont {Storkey}},\ }in\ \href
  {http://papers.nips.cc/paper/5978-covariance-controlled-adaptive-langevin-thermostat-for-large-scale-bayesian-sampling.pdf}
  {\emph {\bibinfo {booktitle} {Advances in Neural Information Processing
  Systems 28}}}\ (\bibinfo  {publisher} {Curran Associates, Inc.},\ \bibinfo
  {year} {2015})\ pp.\ \bibinfo {pages} {37--45}\BibitemShut {NoStop}%
\bibitem [{\citenamefont {Noid}(2013)}]{noid2013perspective}%
  \BibitemOpen
  \bibfield  {author} {\bibinfo {author} {\bibfnamefont {W.}~\bibnamefont
  {Noid}},\ }\href@noop {} {\bibfield  {journal} {\bibinfo  {journal} {J. Chem.
  Phys.}\ }\textbf {\bibinfo {volume} {139}},\ \bibinfo {pages} {090901}
  (\bibinfo {year} {2013})}\BibitemShut {NoStop}%
\bibitem [{\citenamefont {Fukuda}\ and\ \citenamefont
  {Moritsugu}(2017)}]{Fukuda2017}%
  \BibitemOpen
  \bibfield  {author} {\bibinfo {author} {\bibfnamefont {I.}~\bibnamefont
  {Fukuda}}\ and\ \bibinfo {author} {\bibfnamefont {K.}~\bibnamefont
  {Moritsugu}},\ }\href {http://stacks.iop.org/1751-8121/50/i=1/a=015002}
  {\bibfield  {journal} {\bibinfo  {journal} {J. Phys. A: Math. Theor.}\
  }\textbf {\bibinfo {volume} {50}},\ \bibinfo {pages} {015002} (\bibinfo
  {year} {2017})}\BibitemShut {NoStop}%
\bibitem [{\citenamefont {Fukuda}\ and\ \citenamefont
  {Moritsugu}(2015)}]{FukudaMoritsugu2015}%
  \BibitemOpen
  \bibfield  {author} {\bibinfo {author} {\bibfnamefont {I.}~\bibnamefont
  {Fukuda}}\ and\ \bibinfo {author} {\bibfnamefont {K.}~\bibnamefont
  {Moritsugu}},\ }\href {\doibase 10.1088/1751-8113/48/45/455001} {\bibfield
  {journal} {\bibinfo  {journal} {J. Phys. A: Math. Theor.}\ }\textbf {\bibinfo
  {volume} {48}},\ \bibinfo {pages} {455001} (\bibinfo {year}
  {2015})}\BibitemShut {NoStop}%
\bibitem [{\citenamefont {Leimkuhler}\ and\ \citenamefont
  {Matthews}(2013)}]{LeimkuhlerMatthews2013}%
  \BibitemOpen
  \bibfield  {author} {\bibinfo {author} {\bibfnamefont {B.}~\bibnamefont
  {Leimkuhler}}\ and\ \bibinfo {author} {\bibfnamefont {C.}~\bibnamefont
  {Matthews}},\ }\href {\doibase 10.1063/1.4802990} {\bibfield  {journal}
  {\bibinfo  {journal} {J. Chem. Phys.}\ }\textbf {\bibinfo {volume} {138}},\
  \bibinfo {pages} {174102} (\bibinfo {year} {2013})}\BibitemShut {NoStop}%
\bibitem [{\citenamefont {Hoover}(2012)}]{hoover2012computational}%
  \BibitemOpen
  \bibfield  {author} {\bibinfo {author} {\bibfnamefont {W.~G.}\ \bibnamefont
  {Hoover}},\ }\href@noop {} {\emph {\bibinfo {title} {Computational
  statistical mechanics}}}\ (\bibinfo  {publisher} {Elsevier},\ \bibinfo {year}
  {2012})\BibitemShut {NoStop}%
\bibitem [{\citenamefont {Hoover}(1985)}]{hoover1985canonical}%
  \BibitemOpen
  \bibfield  {author} {\bibinfo {author} {\bibfnamefont {W.~G.}\ \bibnamefont
  {Hoover}},\ }\href@noop {} {\bibfield  {journal} {\bibinfo  {journal} {Phys.
  Rev. A}\ }\textbf {\bibinfo {volume} {31}},\ \bibinfo {pages} {1695}
  (\bibinfo {year} {1985})}\BibitemShut {NoStop}%
\bibitem [{\citenamefont {Nos{\'e}}(1984)}]{nose1984molecular}%
  \BibitemOpen
  \bibfield  {author} {\bibinfo {author} {\bibfnamefont {S.}~\bibnamefont
  {Nos{\'e}}},\ }\href@noop {} {\bibfield  {journal} {\bibinfo  {journal} {Mol.
  Phys.}\ }\textbf {\bibinfo {volume} {52}},\ \bibinfo {pages} {255} (\bibinfo
  {year} {1984})}\BibitemShut {NoStop}%
\bibitem [{\citenamefont {{Bussi}}\ \emph {et~al.}(2009)\citenamefont
  {{Bussi}}, \citenamefont {{Zykova-Timan}},\ and\ \citenamefont
  {{Parrinello}}}]{BussiZykova-TimanParrinello2009}%
  \BibitemOpen
  \bibfield  {author} {\bibinfo {author} {\bibfnamefont {G.}~\bibnamefont
  {{Bussi}}}, \bibinfo {author} {\bibfnamefont {T.}~\bibnamefont
  {{Zykova-Timan}}}, \ and\ \bibinfo {author} {\bibfnamefont {M.}~\bibnamefont
  {{Parrinello}}},\ }\href {\doibase 10.1063/1.3073889} {\bibfield  {journal}
  {\bibinfo  {journal} {J. Chem. Phys.}\ }\textbf {\bibinfo {volume} {130}},\
  \bibinfo {pages} {074101} (\bibinfo {year} {2009})}\BibitemShut {NoStop}%
\bibitem [{\citenamefont {Ruelle}(2004)}]{Ruelle2004}%
  \BibitemOpen
  \bibfield  {author} {\bibinfo {author} {\bibfnamefont {D.}~\bibnamefont
  {Ruelle}},\ }\href@noop {} {\bibfield  {journal} {\bibinfo  {journal} {Phys.
  Today}\ }\textbf {\bibinfo {volume} {57}},\ \bibinfo {pages} {48} (\bibinfo
  {year} {2004})}\BibitemShut {NoStop}%
\bibitem [{\citenamefont {Rugh}(1997)}]{rugh1997dynamical}%
  \BibitemOpen
  \bibfield  {author} {\bibinfo {author} {\bibfnamefont {H.~H.}\ \bibnamefont
  {Rugh}},\ }\href@noop {} {\bibfield  {journal} {\bibinfo  {journal} {Phys.
  Rev. Lett.}\ }\textbf {\bibinfo {volume} {78}},\ \bibinfo {pages} {772}
  (\bibinfo {year} {1997})}\BibitemShut {NoStop}%
\bibitem [{\citenamefont {Jepps}\ \emph {et~al.}(2000)\citenamefont {Jepps},
  \citenamefont {Ayton},\ and\ \citenamefont {Evans}}]{jepps2000microscopic}%
  \BibitemOpen
  \bibfield  {author} {\bibinfo {author} {\bibfnamefont {O.~G.}\ \bibnamefont
  {Jepps}}, \bibinfo {author} {\bibfnamefont {G.}~\bibnamefont {Ayton}}, \ and\
  \bibinfo {author} {\bibfnamefont {D.~J.}\ \bibnamefont {Evans}},\ }\href@noop
  {} {\bibfield  {journal} {\bibinfo  {journal} {Phys. Rev. E}\ }\textbf
  {\bibinfo {volume} {62}},\ \bibinfo {pages} {4757} (\bibinfo {year}
  {2000})}\BibitemShut {NoStop}%
\bibitem [{\citenamefont {{Hoover}}\ and\ \citenamefont
  {{Holian}}(1996)}]{HooverHolian1996}%
  \BibitemOpen
  \bibfield  {author} {\bibinfo {author} {\bibfnamefont {W.~G.}\ \bibnamefont
  {{Hoover}}}\ and\ \bibinfo {author} {\bibfnamefont {B.~L.}\ \bibnamefont
  {{Holian}}},\ }\href {\doibase 10.1016/0375-9601(95)00973-6} {\bibfield
  {journal} {\bibinfo  {journal} {Phys Lett A}\ }\textbf {\bibinfo {volume}
  {211}},\ \bibinfo {pages} {253} (\bibinfo {year} {1996})}\BibitemShut
  {NoStop}%
\bibitem [{\citenamefont {Uhlenbeck}\ and\ \citenamefont
  {Ford}(1963)}]{uhlenbeck1963lectures}%
  \BibitemOpen
  \bibfield  {author} {\bibinfo {author} {\bibfnamefont {G.}~\bibnamefont
  {Uhlenbeck}}\ and\ \bibinfo {author} {\bibfnamefont {G.}~\bibnamefont
  {Ford}},\ }\href@noop {} {\emph {\bibinfo {title} {Lectures in statistical
  mechanics.}}}\ (\bibinfo  {publisher} {AMS, Providence, Rhode Island},\
  \bibinfo {year} {1963})\BibitemShut {NoStop}%
\bibitem [{\citenamefont {Novikov}(1965)}]{novikov1965functionals}%
  \BibitemOpen
  \bibfield  {author} {\bibinfo {author} {\bibfnamefont {E.~A.}\ \bibnamefont
  {Novikov}},\ }\href@noop {} {\bibfield  {journal} {\bibinfo  {journal}
  {Soviet Physics-JETP}\ }\textbf {\bibinfo {volume} {20}},\ \bibinfo {pages}
  {1290} (\bibinfo {year} {1965})}\BibitemShut {NoStop}%
\bibitem [{\citenamefont {Klyatskin}(2005)}]{klyatskin2005dynamics}%
  \BibitemOpen
  \bibfield  {author} {\bibinfo {author} {\bibfnamefont {V.~I.}\ \bibnamefont
  {Klyatskin}},\ }\href@noop {} {\emph {\bibinfo {title} {Dynamics of
  stochastic systems}}}\ (\bibinfo  {publisher} {Elsevier},\ \bibinfo {year}
  {2005})\BibitemShut {NoStop}%
\bibitem [{\citenamefont {Mattingly}\ \emph {et~al.}(2002)\citenamefont
  {Mattingly}, \citenamefont {Stuart},\ and\ \citenamefont
  {Higham}}]{mattingly2002ergodicity}%
  \BibitemOpen
  \bibfield  {author} {\bibinfo {author} {\bibfnamefont {J.~C.}\ \bibnamefont
  {Mattingly}}, \bibinfo {author} {\bibfnamefont {A.~M.}\ \bibnamefont
  {Stuart}}, \ and\ \bibinfo {author} {\bibfnamefont {D.~J.}\ \bibnamefont
  {Higham}},\ }\href@noop {} {\bibfield  {journal} {\bibinfo  {journal} {Stoch.
  Proc. Appl.}\ }\textbf {\bibinfo {volume} {101}},\ \bibinfo {pages} {185}
  (\bibinfo {year} {2002})}\BibitemShut {NoStop}%
\bibitem [{\citenamefont {Samoletov}(1999)}]{Samoletov1999}%
  \BibitemOpen
  \bibfield  {author} {\bibinfo {author} {\bibfnamefont {A.~A.}\ \bibnamefont
  {Samoletov}},\ }\href {\doibase 10.1023/A:1004656820908} {\bibfield
  {journal} {\bibinfo  {journal} {J. Stat. Phys.}\ }\textbf {\bibinfo {volume}
  {96}},\ \bibinfo {pages} {1351} (\bibinfo {year} {1999})}\BibitemShut
  {NoStop}%
\bibitem [{\citenamefont {{Hoover}}\ \emph {et~al.}(2015)\citenamefont
  {{Hoover}}, \citenamefont {{Sprott}},\ and\ \citenamefont
  {{Patra}}}]{HooverSprottPatra2015}%
  \BibitemOpen
  \bibfield  {author} {\bibinfo {author} {\bibfnamefont {W.~G.}\ \bibnamefont
  {{Hoover}}}, \bibinfo {author} {\bibfnamefont {J.~C.}\ \bibnamefont
  {{Sprott}}}, \ and\ \bibinfo {author} {\bibfnamefont {P.~K.}\ \bibnamefont
  {{Patra}}},\ }\href {\doibase 10.1016/j.physleta.2015.08.034} {\bibfield
  {journal} {\bibinfo  {journal} {Phys Lett A}\ }\textbf {\bibinfo {volume}
  {379}},\ \bibinfo {pages} {2935} (\bibinfo {year} {2015})},\ \Eprint
  {http://arxiv.org/abs/1503.06749} {arXiv:1503.06749 [cond-mat.stat-mech]}
  \BibitemShut {NoStop}%
\bibitem [{\citenamefont {Hoover}\ \emph {et~al.}(2016)\citenamefont {Hoover},
  \citenamefont {Sprott},\ and\ \citenamefont {Hoover}}]{Hoover2016}%
  \BibitemOpen
  \bibfield  {author} {\bibinfo {author} {\bibfnamefont {W.~G.}\ \bibnamefont
  {Hoover}}, \bibinfo {author} {\bibfnamefont {J.~C.}\ \bibnamefont {Sprott}},
  \ and\ \bibinfo {author} {\bibfnamefont {C.~G.}\ \bibnamefont {Hoover}},\
  }\href@noop {} {\bibfield  {journal} {\bibinfo  {journal} {Commun. Nonlinear
  Sci. Numer. Simul.}\ }\textbf {\bibinfo {volume} {32}},\ \bibinfo {pages}
  {234} (\bibinfo {year} {2016})}\BibitemShut {NoStop}%
\bibitem [{\citenamefont {{Patra}}\ and\ \citenamefont
  {{Bhattacharya}}(2014)}]{PatraBhattacharya2014}%
  \BibitemOpen
  \bibfield  {author} {\bibinfo {author} {\bibfnamefont {P.~K.}\ \bibnamefont
  {{Patra}}}\ and\ \bibinfo {author} {\bibfnamefont {B.}~\bibnamefont
  {{Bhattacharya}}},\ }\href {\doibase 10.1063/1.4864204} {\bibfield  {journal}
  {\bibinfo  {journal} {J. Chem. Phys.}\ }\textbf {\bibinfo {volume} {140}},\
  \bibinfo {eid} {064106} (\bibinfo {year} {2014})}\BibitemShut {NoStop}%
\bibitem [{\citenamefont {Patra}\ and\ \citenamefont
  {Bhattacharya}(2015)}]{patra2015ergodic}%
  \BibitemOpen
  \bibfield  {author} {\bibinfo {author} {\bibfnamefont {P.~K.}\ \bibnamefont
  {Patra}}\ and\ \bibinfo {author} {\bibfnamefont {B.}~\bibnamefont
  {Bhattacharya}},\ }\href@noop {} {\bibfield  {journal} {\bibinfo  {journal}
  {J. Chem. Phys.}\ }\textbf {\bibinfo {volume} {142}},\ \bibinfo {pages}
  {194103} (\bibinfo {year} {2015})}\BibitemShut {NoStop}%
\bibitem [{\citenamefont {Legoll}\ \emph {et~al.}(2007)\citenamefont {Legoll},
  \citenamefont {Luskin},\ and\ \citenamefont {Moeckel}}]{legoll2007non}%
  \BibitemOpen
  \bibfield  {author} {\bibinfo {author} {\bibfnamefont {F.}~\bibnamefont
  {Legoll}}, \bibinfo {author} {\bibfnamefont {M.}~\bibnamefont {Luskin}}, \
  and\ \bibinfo {author} {\bibfnamefont {R.}~\bibnamefont {Moeckel}},\
  }\href@noop {} {\bibfield  {journal} {\bibinfo  {journal} {Arch. Ration.
  Mech. Anal.}\ }\textbf {\bibinfo {volume} {184}},\ \bibinfo {pages} {449}
  (\bibinfo {year} {2007})}\BibitemShut {NoStop}%
\bibitem [{\citenamefont {Martyna}\ \emph {et~al.}(1992)\citenamefont
  {Martyna}, \citenamefont {Klein},\ and\ \citenamefont
  {Tuckerman}}]{martyna1992nose}%
  \BibitemOpen
  \bibfield  {author} {\bibinfo {author} {\bibfnamefont {G.~J.}\ \bibnamefont
  {Martyna}}, \bibinfo {author} {\bibfnamefont {M.~L.}\ \bibnamefont {Klein}},
  \ and\ \bibinfo {author} {\bibfnamefont {M.}~\bibnamefont {Tuckerman}},\
  }\href@noop {} {\bibfield  {journal} {\bibinfo  {journal} {The Journal of
  chemical physics}\ }\textbf {\bibinfo {volume} {97}},\ \bibinfo {pages}
  {2635} (\bibinfo {year} {1992})}\BibitemShut {NoStop}%
\end{thebibliography}%

\end{document}